\documentclass[prb,twocolumn,showpacs,byrevtex]{revtex4}
\usepackage{epsf}
\usepackage{graphicx}
\begin{document}

\title{{\it Ab initio} calculations of structural and electronic properties of CdTe clusters}

\author{Somesh Kr. Bhattacharya} \email{skb@physics.unipune.ernet.in}

\author{Anjali Kshirsagar$^1$} \email{anjali@physics.unipune.ernet.in}

\affiliation {Department of Physics, 
              University of Pune, 
              Pune 411 007, India.}

\date{\today}

\begin{abstract}
We present results of a study of small stoichiometric $Cd_nTe_n$ 
($1{\leq}n{\leq}6$)
clusters and few medium sized non-stoichiometric $Cd_mTe_n$
[($m,n~=~13,~16,~19$);~($m{\neq}n$)] clusters
using the Density Functional formalism and projector augmented wave method 
within the generalized gradient approximation. Structural properties
{\it viz.} geometry, bond length, symmetry and electronic properties
like HOMO-LUMO gap, binding energy, ionization potential and nature
of bonding {\it etc.} have been analyzed.
Medium sized non-stoichiometric clusters were considered
as fragments of the bulk with T{$_{d}$} symmetry. It was observed that
upon relaxation, the symmetry changes for the Cd-rich clusters
whereas the Te-rich clusters retain their symmetry. The Cd-rich clusters
develop a HOMO-LUMO gap due to relaxation whereas there
is no change in the HOMO-LUMO gap of the Te-rich clusters.
Thus, the symmetry of a cluster seems to be an important factor 
in determining the HOMO-LUMO gap.
\end{abstract}


\maketitle

\begin{section}{Introduction}
Semiconductor nanoparticles or Quantum Dots (QDs),
in particular of II-VI materials, have received tremendous
attention during the last two decades owing to their unusual physical properties
and wide range of applications~\cite{r1}. A systematic study of QDs is useful 
to understand the evolution of their physical and chemical properties with 
size. Moving from a molecule to bulk, one can observe a range of variation of 
fundamental properties. This size
dependent effect, however, is most significant for small sized clusters and is 
attributed to the different geometry of clusters in comparision to the 
bulk. Quantum confinement
of electronic states and large surface to volume ratio modify the
properties of these clusters. In bare clusters, unsaturated bonds lead to 
peculiar structural and electronic properties like surface reconstruction, 
formation of new cleavage planes etc.

The differences in the structural properties also affect the electronic 
properties of the QDs. In bulk systems one observes quasi-continuous 
electronic levels forming bands
whereas in clusters the electronic levels are discrete. Apart
from the structural and electronic properties, other properties like the 
thermodynamic
and optical properties too differ significantly for clusters as compared to
bulk~\cite{r2,r3}.

Among the II - VI semiconductors, CdX (X = S, Se, Te)
have potential application in optical 
and optoelectronic devices. 
CdTe nanoclusters, in particular, are used in optical sensors.
Multilayered CdTe with poly (diallyl dimethyl ammonium 
chloride) are electro and photo-active and hence are used in
neuroprosthetic devices~\cite{pdda}. Self assembled CdTe nanodots have high 
efficiency in terms
of photoluminescence and hence are very good for the purpose of making LEDs.
CdTe and Au nanohybrid materials have photoluminescence variation depending upon
environmental condition like temperature, ionic strength, solvents and hence are
used as biosensors~\cite{au-cdte}. 

We present Density Functional Theory (DFT)
based calculations for some small, Cd$_n$Te$_n$ ($1{\leq}n{\leq}6$),
and few medium sized Cd$_m$Te$_n$ ($m,n=13,16,19$)
($m{\neq}n$) clusters.
Both Cd and Te in bulk phase exist in hexagonal structure
with c/a ratio 1.89 and 1.33 
respectively~\cite{ash} whereas CdTe in bulk phase has lattice consatnt
$a=6.48\AA$ in zincblende structure and $a=4.57\AA$ and $c=7.47\AA$ in
wurtzite structure.
CdTe has a calculated direct band gap of 1.59eV at the 
$\Gamma$ point~\cite{che}.

Study of small clusters is challenging due to lack of experimental structural 
information and increased degree of freedom. Very small clusters (containing 
up to 8-10 atoms) exhibit symmetry and have only a few 
possible geometries. On the other hand, medium-sized clusters can be considered as fragments of bulk possessing the same structural symmetry as that of the 
bulk~\cite{r4}. To the best of our knowledge, so far no {\it ab-initio} 
calculations have been reported on CdTe QDs.
We compare our results for small CdTe QDs with those of CdS and CdSe 
wherever available.

Structural, electronic and optical properties
of II-VI semiconductor nanoclusters have been studied using
a variety of methods 
based on {\it ab-initio} DFT, quantum chemistry methods like Hartree-
Fock, Gaussian etc. and approximate methods
like tight-binding method~\cite{dft}-\cite{tb}. 

\end{section}

\begin{section}{Methodology and Computational Details}
The first step for calculating the cluster properties is the determination
of the lowest energy configuration. 
Existence 
of several configurations with varying coordination and multiple minima 
in the potential energy
surface enormously complicate the problem making it almost impossible to 
decide by direct calculation of the lowest energy configuration for clusters.

In this work we have used Born-Oppenheimer 
molecular dynamics within the Kohn-Sham
density functional framework~\cite{Hon}. The electronic structure 
is calculated self-consistently using projector augmented wave
(PAW) method~\cite{Blo} as implemented in VASP package~\cite{vasp} 
within the framework of GGA instead of{\it Vanderbilt} ultra-soft 
pseudo-potential(US-PP)~\cite{Van} as it provides better first principle 
simulations as compared to US-PP~\cite{Blo}.
In pseudo potential methods, the effect of core electrons and 
nuclei is replaced by an effective ionic potential and only the 
valence electrons which are directly involved in chemical bonding 
are considered.

Various parameterizations have been 
used for different exchange-correlation potentials~\cite{exch}
in the DFT formalism in the literature. 
It has been reported that generalized gradient approximation (GGA) gives much
better results for binding energy and band gap
as compared to local density approximation (LDA) and
therefore we have used GGA, as developed by Perdew-Berke-Ernzerhof 
(PBE)~\cite{pbe} for our calculation.
PBE is known to improve total energy and  molecule atomization energy
and therefore the electronic properties,
as compared to PW91~\cite{pw91} and the results agree with that of the 
experiments very
well. 

The geometry of small clusters varies significantly from 
that of the bulk. We have chosen several possible initial
geometries by exploiting the symmetry considerations
and by exchanging the positions of anionic and cationic atoms.
The number of initial structures 
is small for very small clusters. But for clusters with 
10-12 atoms, lot many configurations have been used to determine 
the lowest energy structure.  

For structure optimization we have used a super-cell of size $a=25$\AA\ and
the plane wave cut-off was set to be 274.3 eV. These
values are chosen after performing calculations 
for different cell sizes and cutoff energies.
We have employed conjugate gradient (CG) technique implemented via Kosugi
algorithm~\cite{kos} (special Davison block iteration scheme) for 
optimizing geometry.

The valence electron configurations used
for Cd and Te are $5s^24d^{10}$  and $5s^25p^4$ respectively.
The $4d$ levels in Te atom are well separated from
the $5s$ level and hence can be included
in the core. 

CdTe in bulk has zincblende (ZB) structure with T{$_d$} symmetry. 
Therefore for medium sized non-stoichiometric
clusters Cd$_m$Te$_n$ ($m{\neq}n$), the initial
geometry was considered as a fragment of the ZB structure.
The size of the super-cell was chosen to be $a=30\AA$ which is
sufficiently large to minimize
inter-cluster interactions even for the largest cluster
size we consider in this work.
In all the structure minimizations, the force and energy
convergence obtained was $\sim 10^{-3}$ eV/$\AA$ 
and $\sim 10^{-4}$ eV respectively. 
\end{section}

\begin{section}{Results and Discussions}

\begin{subsection}{Structural properties of Cd$_n$Te$_n$}
The geometries for lowest energy (LE) of
Cd$_n$Te$_n$ obtained by CG are shown in
Fig. \ref{geo1} and the first local minima (FLM) in Fig. \ref{geo2}.
It is interesting to note that for all the calculated lowest energy structures, 
the chalcogenide atoms are at the peripheral positions. This
 is a direct consequence of 
Coulomb repulsion between the lone-pairs situated on
 the chalcogenide atoms which is minimized 
by separating the chalcogenide atoms towards the peripheral position.
As expected, the ground state structures of these CdTe clusters
do not resemble the bulk phase.

\begin{figure}
\epsfxsize 1.0in
\epsffile {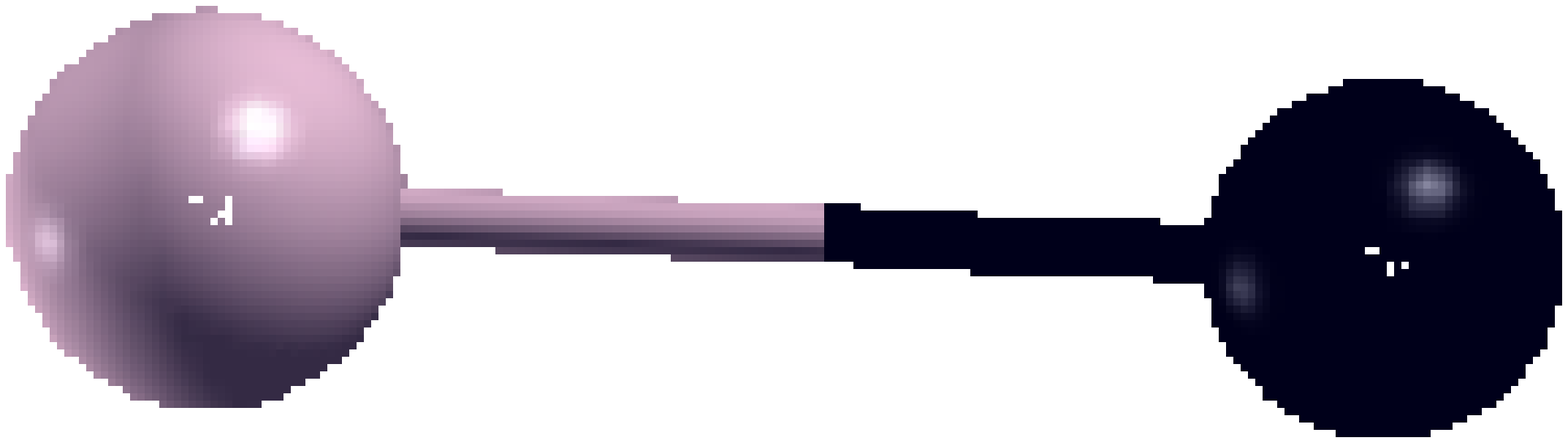}
\epsfxsize 1.1in
\epsffile {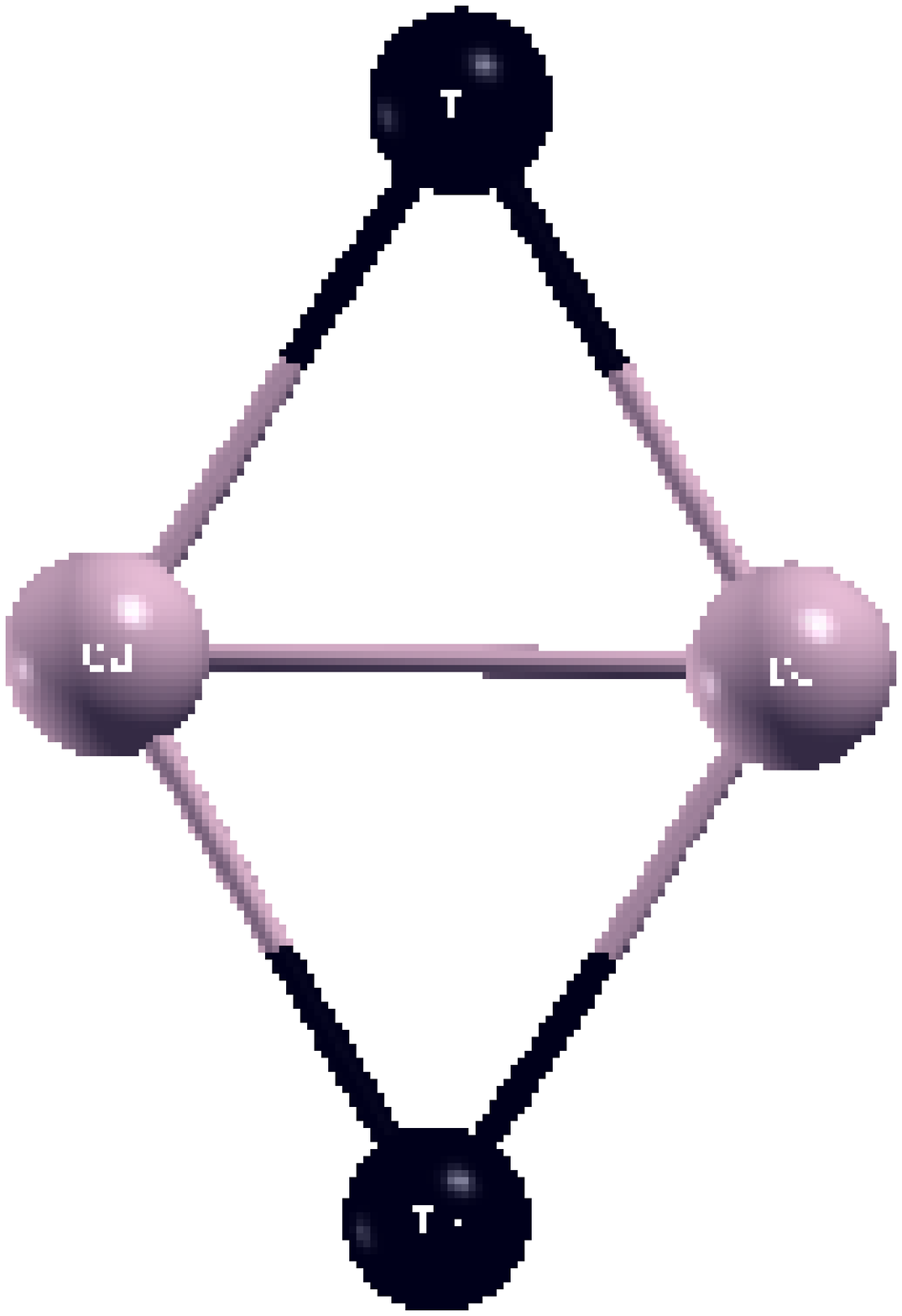}
\vspace{0.1in}
\epsfxsize 1.0in
\epsffile {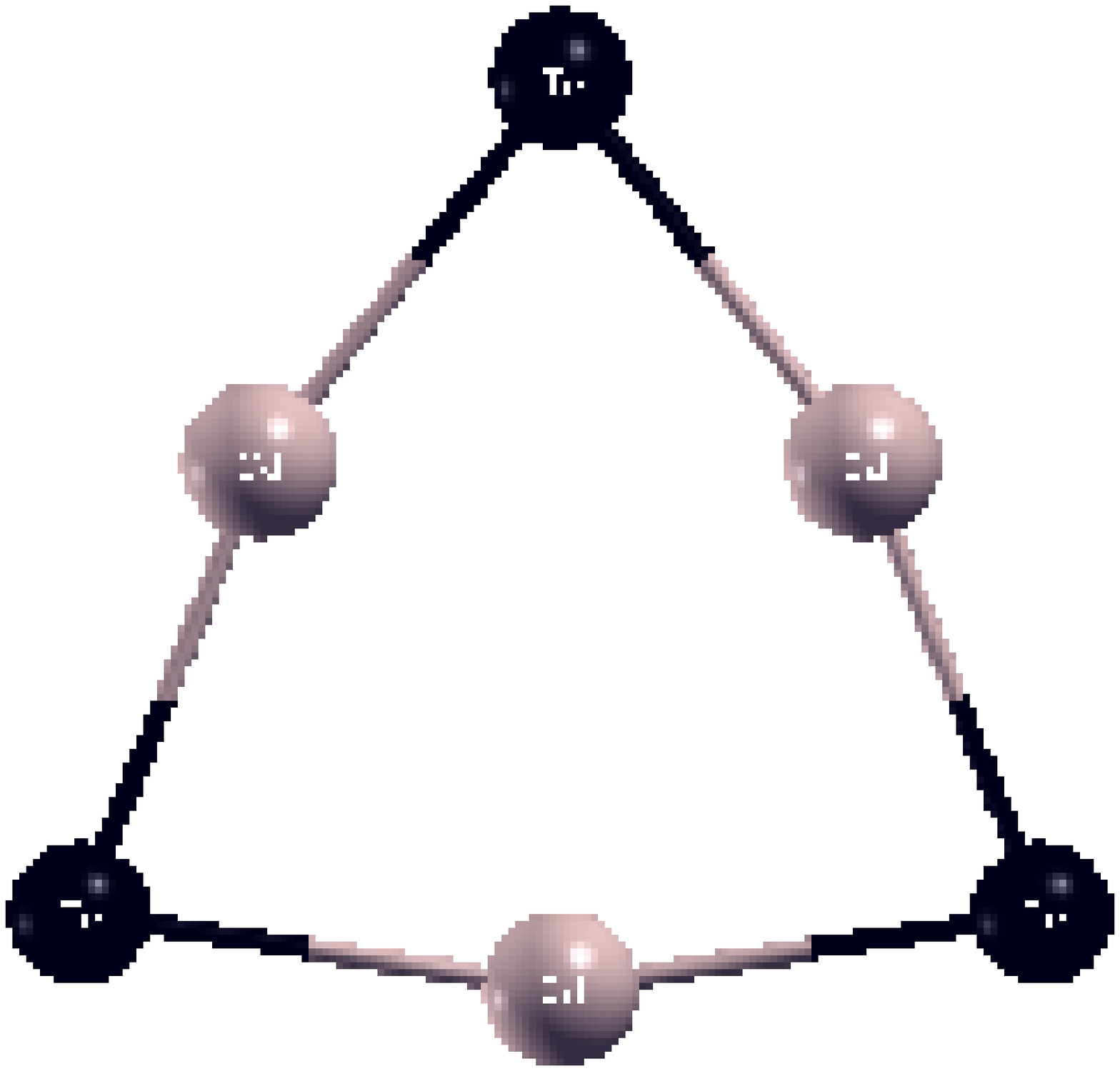}
\epsfxsize 1.3in
\epsffile {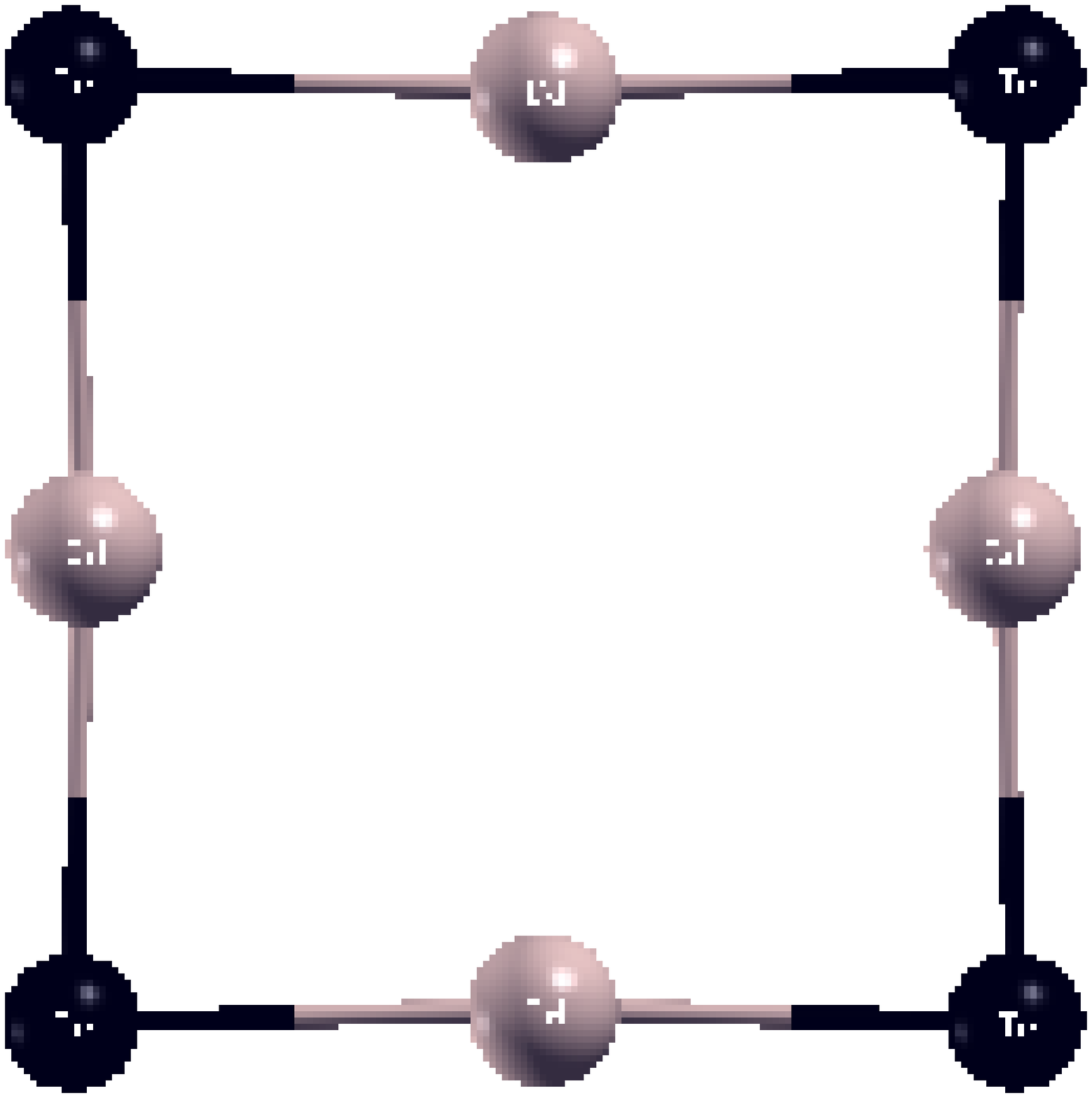}
\epsfxsize 2.0in
\epsffile {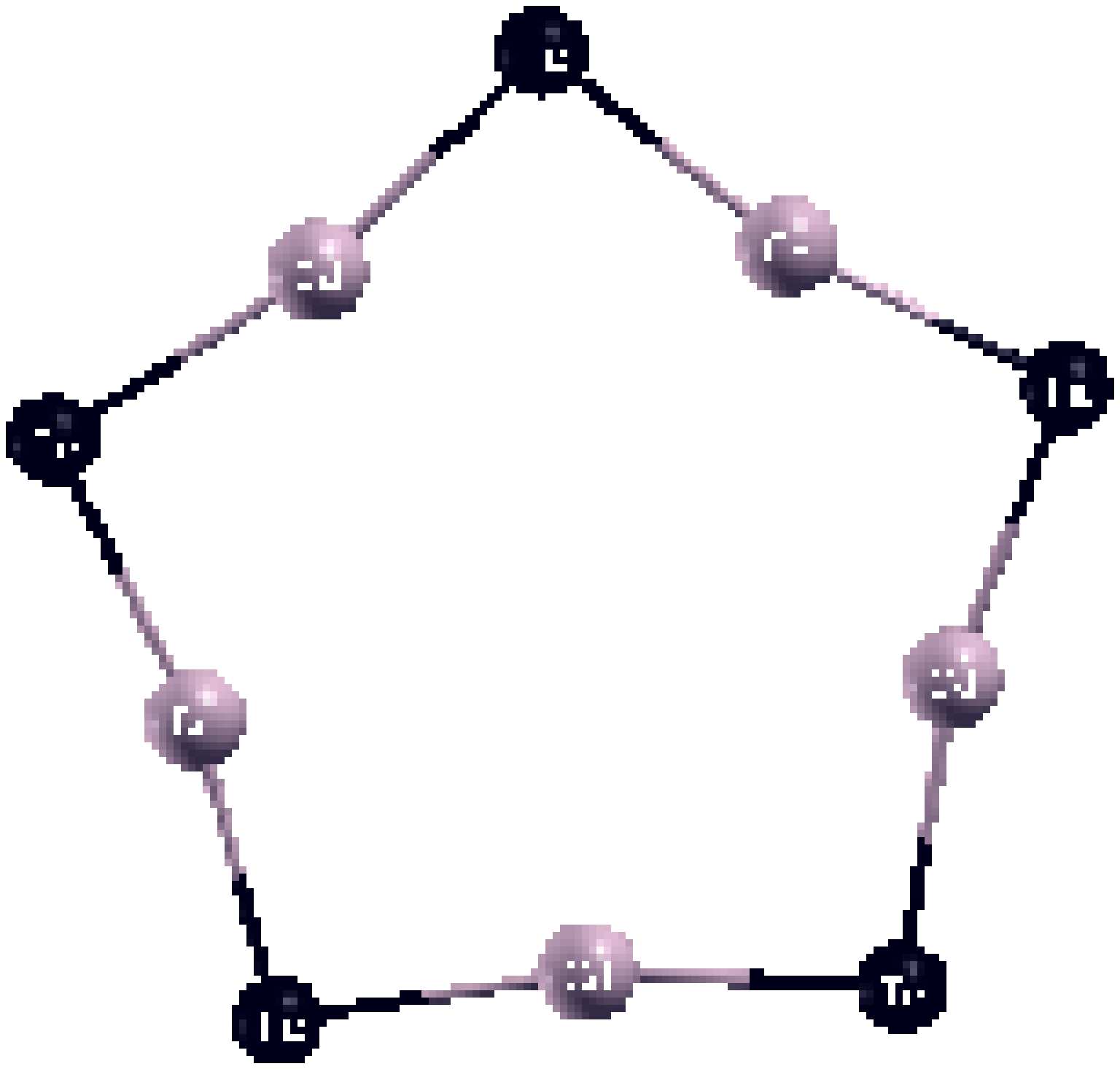}
\epsfxsize 2.5in
\epsffile {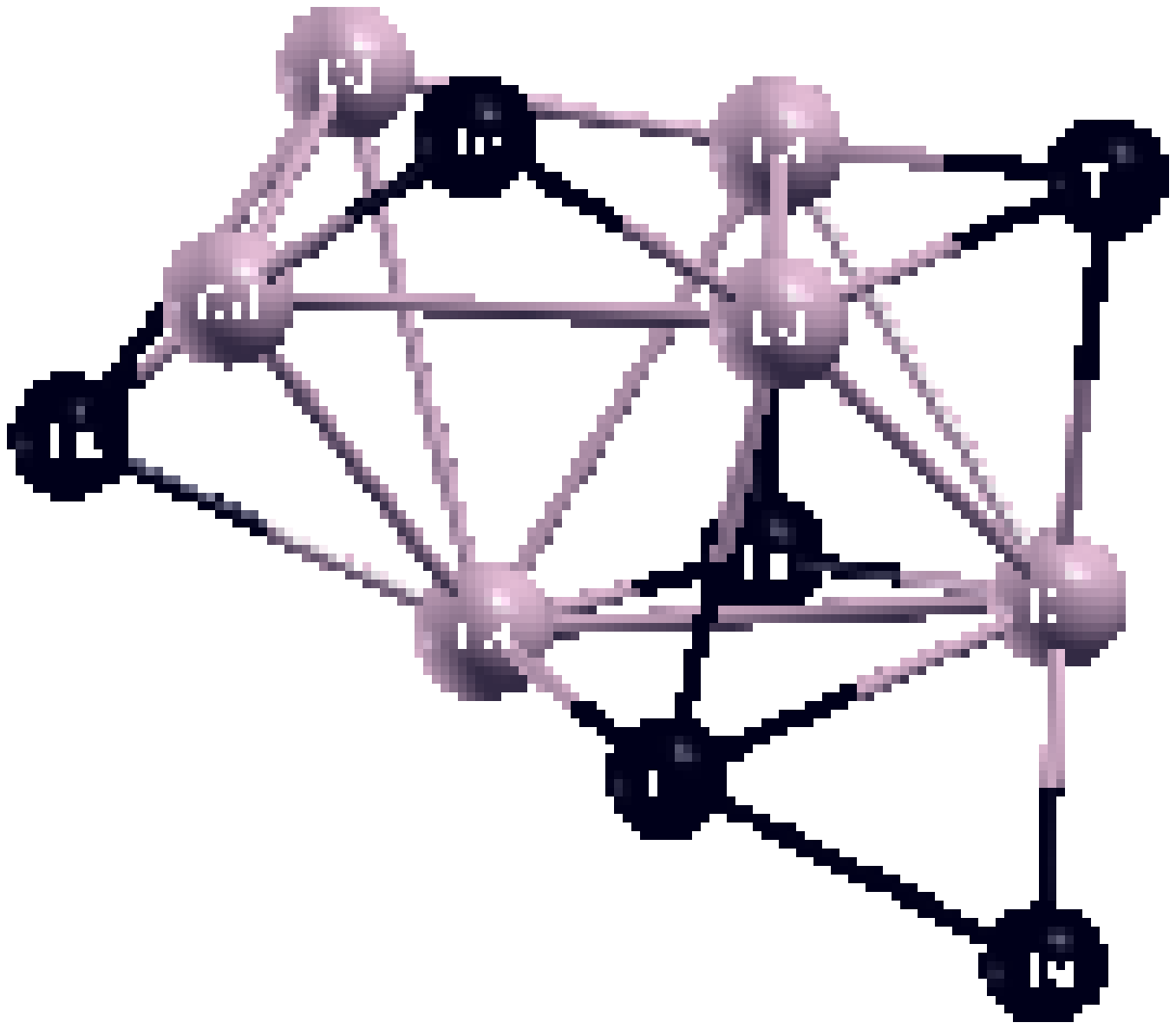}
\vspace{-0.2in}
\caption{Lowest energy configuration
of Cd$_n$Te$_n$ for $(1{\leq}n{\leq}6)$
The grey atoms are Cd and the black atoms are Te.}\label{geo1}
\end{figure}

\begin{figure}
\epsfxsize 1.4in
\epsffile {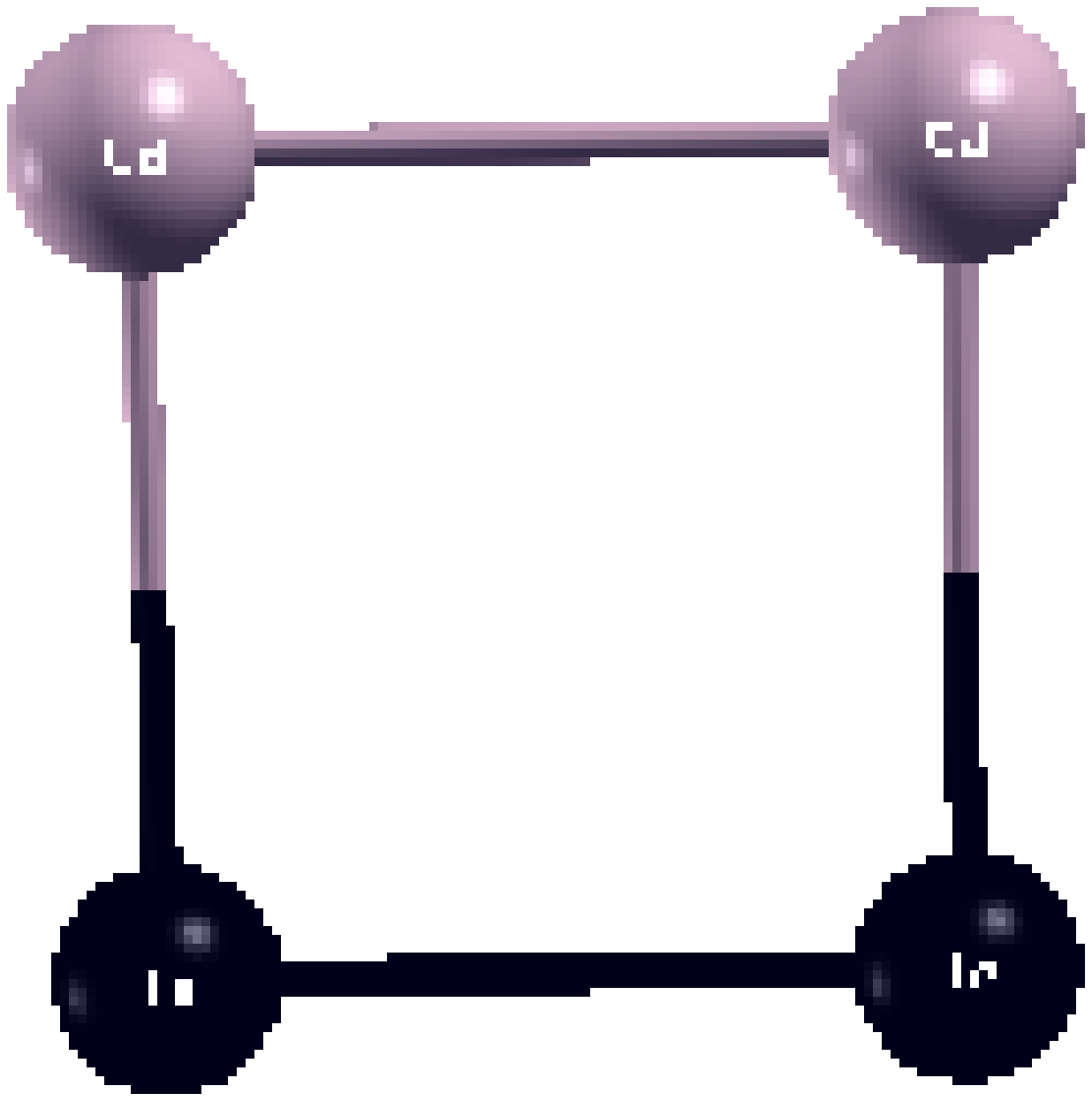}
\epsfxsize 1.6in
\epsffile {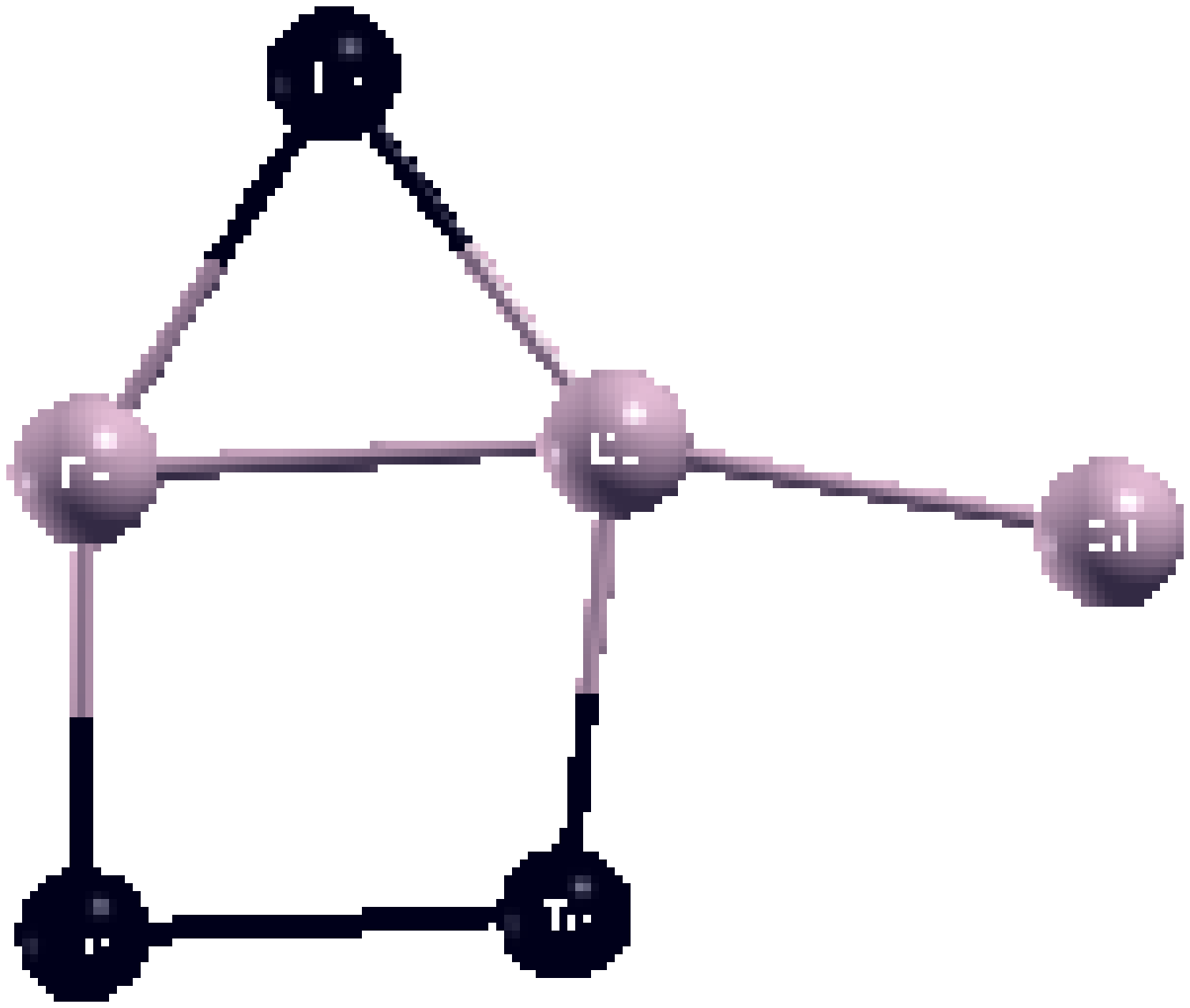}
\epsfxsize 2.5in
\epsffile {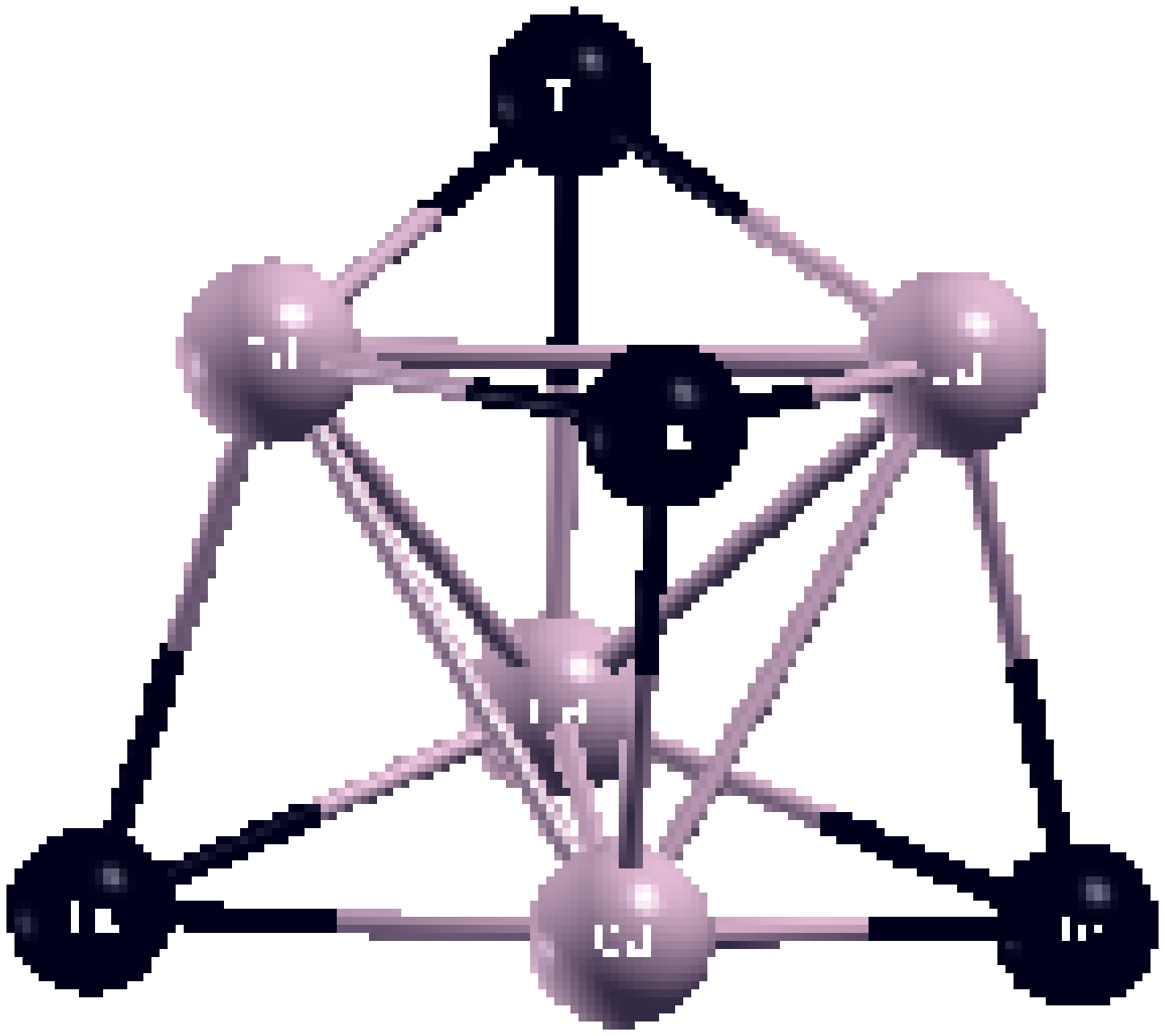}
\epsfxsize 3.0in
\epsffile {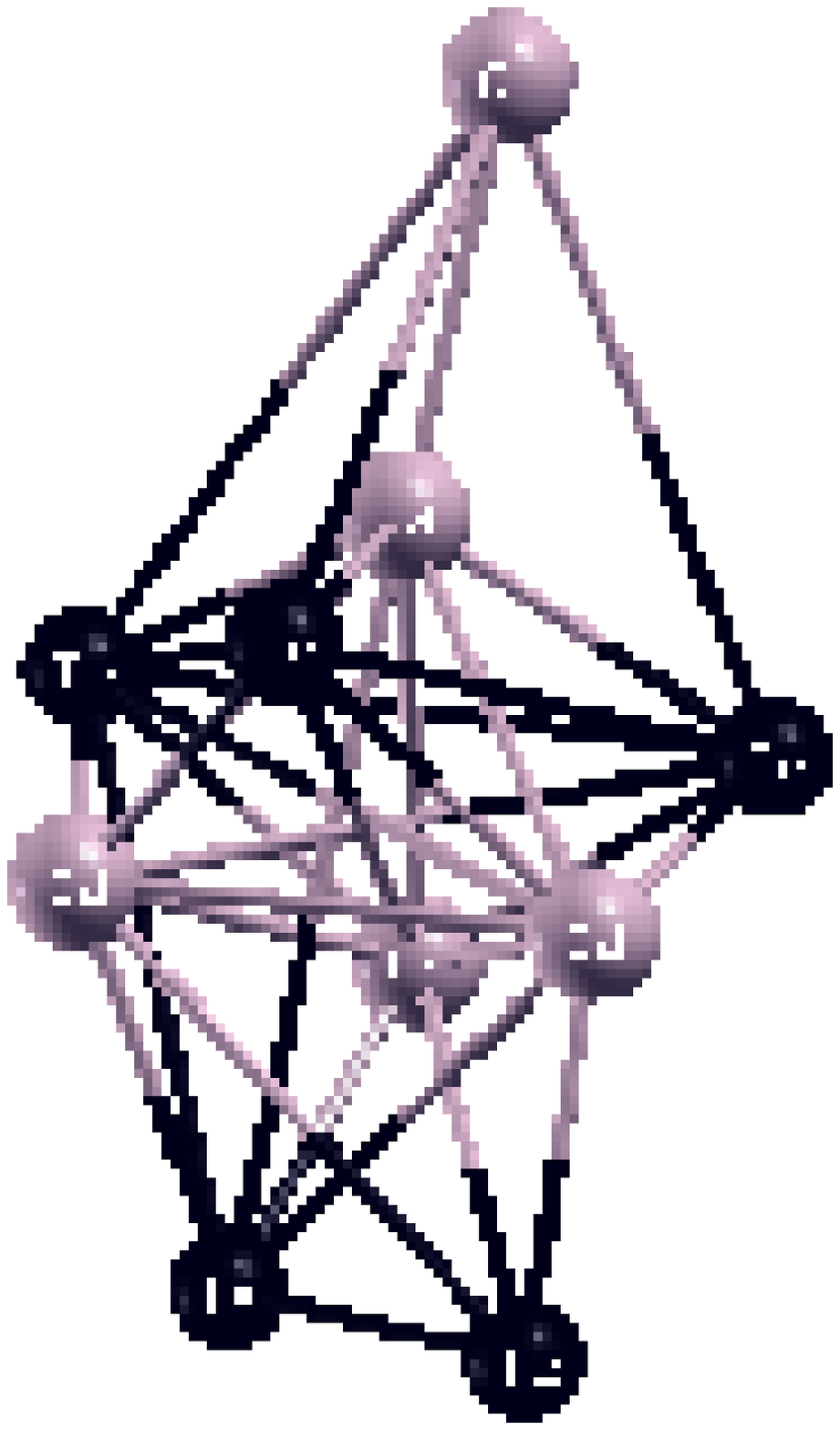}
\epsfxsize 3.0in
\epsffile {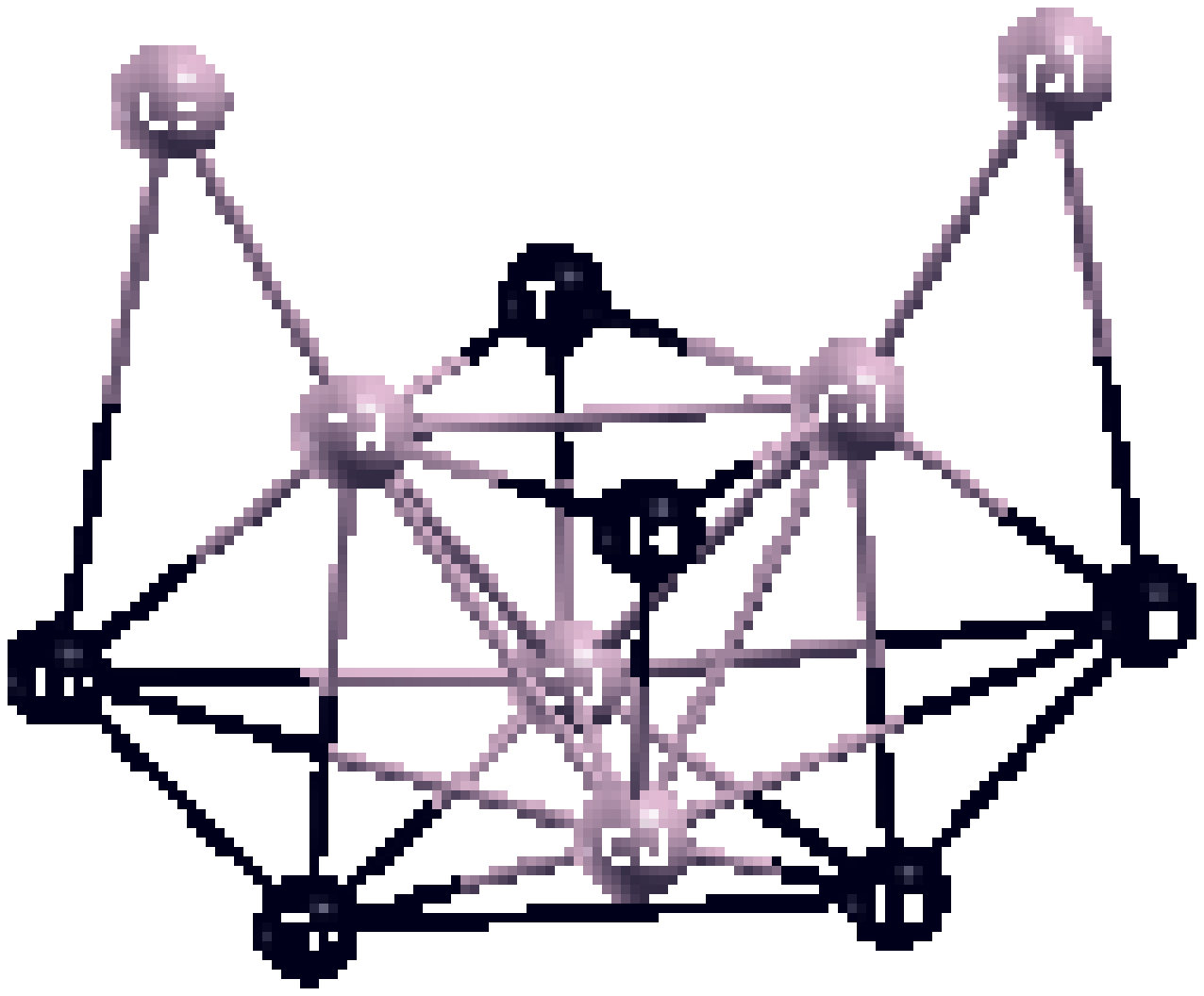}
\vspace {-0.5in}
\caption{Geometries of the first local minima 
of Cd$_n$Te$_n$ for $(1 \leq n \leq 6)$.}\label{geo2} 
\end{figure}

CdTe dimer is a linear molecule with bond length of 2.57$\AA$
and belongs to $C_{\infty v}$ point group.
For Cd$_2$Te$_2$, Cd$_3$Te$_3$,
Cd$_4$Te$_4$, Cd$_5$Te$_5$ and Cd$_6$Te$_6$,
there are two possible lowest energy configurations within
the energy difference of $\sim0.11$eV, $\sim0.22$eV
$\sim0.04$eV, $\sim0.05$eV and $\sim0.09$eV
per atom, respectively. Except for Cd$_6$Te$_6$, the lowest energy 
geometry is a planar ring-like structure.
Cd$_2$Te$_2$ has LE
as the rhombic structure and a square planar structure as 
FLM. The LE and FLM belong to $D_{2h}$ and $C_{2v}$ 
point group repectively. 
Cd$_3$Te$_3$ has triangular structure ($D_{3h}$ point 
group) as LE
configuration and another planar structure ($C_s$ point group) as the FLM.
No three dimensional (3D) stable geometry is obtained for $n = 2, 3$. 
For Cd$_4$Te$_4$, the square planar ($D_{4h}$ point
group) geometry is obtained for LE and 3D 
tetrahedral structure ($T_d$ point group) as FLM. 
 Cd$_5$Te$_5$ is a planar pentagon ($C_s$ point group)
in LE and a non-symmetric 3D structure
as FLM. A transition from planar LE to 3D LE 
occurs  for $n = 6$. This is attributed, as explained later,
to the tendency of achieving higher coordination without
putting too much strain on the bond angles.
All the atoms
in planar clusters, except for the FLM of Cd$_3$Te$_3$, have coordination 2. 

The geometries obtained for the CdTe clusters are quite
similar to those of corresponding CdS and CdSe clusters
upto $n = 4$~\cite{r6}. However,
the geometries of Cd$_5$Te$_5$
and Cd$_6$Te$_6$ are very different from
the CdS and CdSe counterparts~\cite{r6}. This difference may 
be attributed to the fact that we have included 4$d$ of Cd 
in the valence configuration in the present study. The necessity of including 
4$d$ electrons of the cations in the valence configuration
has been emphasized by Wei and Zunger~\cite{Wei}.
 
As we move from planar to 3D structure the coordination of atoms 
increases from 2 to 3 and in some cases 4.
The 3D structures of Cd$_4$Te$_4$, Cd$_5$Te$_5$ and
Cd$_6$Te$_6$ can be considered to be made up of smaller 
cluster units namely Cd$_2$Te$_2$ and
Cd$_3$Te$_3$.

A comparision of Cd$_n$Te$_n$ clusters with Cd$_n$~\cite{cd}
shows that three dimensional 
structures are more favorable in pure Cd clusters as compared to CdTe clusters
where planar structure is favored even for 10 atom cluster.

\begin{figure}[h]
\centering
\includegraphics[height=8.5cm,angle=270] {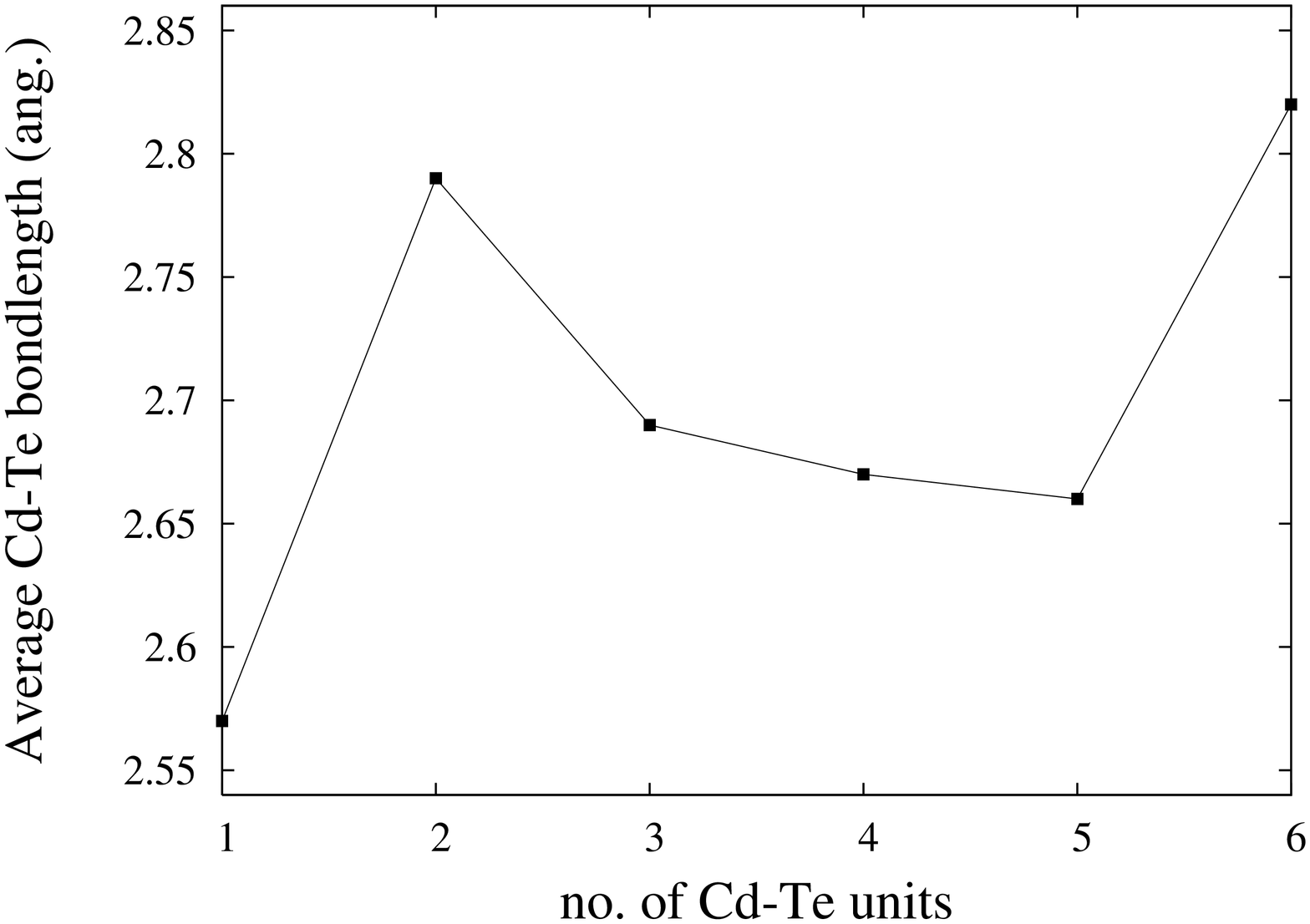}
\caption{Average Cd-Te bond length in $\AA$  {\it vs} number of Cd-Te units.}
\label{av_bl}
\end{figure}

Figure \ref{av_bl} shows the variation of average Cd-Te bond length with the 
number of atoms in the clusters. The near neighbour (nn) 
distance for CdTe bulk is $2.81$\AA. As evident from
Fig. \ref{av_bl}, these clusters have an average bond length
less than the nn distance in bulk. This is due to incomplete coordination 
in these clusters on 
account of very small number of atoms. The bond length of CdTe dimer
is the smallest. In fact, the smallest 
Cd-Te bond length remains almost constant in these clusters for $n = 3-6$ (not 
shown in figure). 
The average Cd-Te bond length shrinks with increasing $n$ in planar structures
($n = 2-5$)
and increases in going from planar to 3D structures.

Surface studies~\cite{vog} on heteropolar covalent and ionic semiconductors
have identified quantum mechanical electron-electron Coulomb repulsion,
hybridization effects and classical Coulomb attraction between anion and cation 
as competing factors in deciding the stability of the surface. The last factor 
seems to be more important with only 3 neighbours for atoms on the surface.
Optimum energy gain is achieved when the distance between anion and cation is 
as small as possible. Thus shorter bonds are more ionic in nature. We therefore 
expect that CdTe dimer will have more ionic character as compared to other 
clusters. In Cd$_2$Te$_2$, there exist a covalent bond between Cd-Cd which 
accounts for an increase in average Cd-Te bond length. As explained later from 
the charge density plots, there are no covalent bonds between either Cd-Cd or
Te-Te in other planar structures ($n = 3-5$). Therefore, the average Cd-Te bond 
length is smaller in these clusters. In Cd$_6$Te$_6$, bonds between similar 
atoms appear and as a result the average Cd-Te bond length increases.

Cd-Cd bond length in Cd$_2$Te$_2$ is smaller (2.87$\AA$) than 
in Cd dimer (3.47 $\AA$). The average Cd-Cd bond length is
same in Cd$_3$Te$_3$ and Cd$_3$. However, it is
larger in Cd$_n$Te$_n$ as compared to Cd$_n$ for
$3 {\leq} n {\leq} 6$. In contrast, average Te-Te bond length 
is significantly larger in CdTe clusters in comparison to 
pure Te clusters as expected due to lone pair repulsion. Also it does not
appreciably change with $n$.

Average Te-Cd-Te bond angle increases from $107.2^o$ in Cd$_2$Te$_2$
to nearly $180^o$ in Cd$_4$Te$_4$. The angle further opens up 
in Cd${_5}$Te${_5}$. In Cd$_6$Te$_6$, all the Te-Cd-Te angles
are smaller than linear angle as a consequence of 3D structure. 
Thus we see that smaller clusters 
have a tendency to form linear Te-Cd-Te bonds resulting in planar 
structures with coordiantion of 2 and as
higher coordination number takes over, a transition from
planar to three dimensional structure results for stable geometry.

All the clusters are found to be stable against fragmentation 
into smaller clusters.
\end{subsection}

\begin{subsection}{Electronic properties of Cd$_n$Te$_n$}
Figures \ref{bg}  and \ref{be} show the variation in the 
highest occupied molecular orbital (HOMO) - lowest unoccupied
molecular orbital (LUMO) gap (E$_g$)
and binding energy (defined in Eq. (\ref{BE})
per Cd-Te unit with the number of Cd-Te units in the cluster.

\vspace{-0.1in}
\begin{equation}
E_b = [ n\{E(Cd)+E(Te)\} - E(Cd_nTe_n) ] / n \label{BE}
\end{equation}

\begin{figure}[h]
\centering
\includegraphics[height=8.5cm,angle=270] {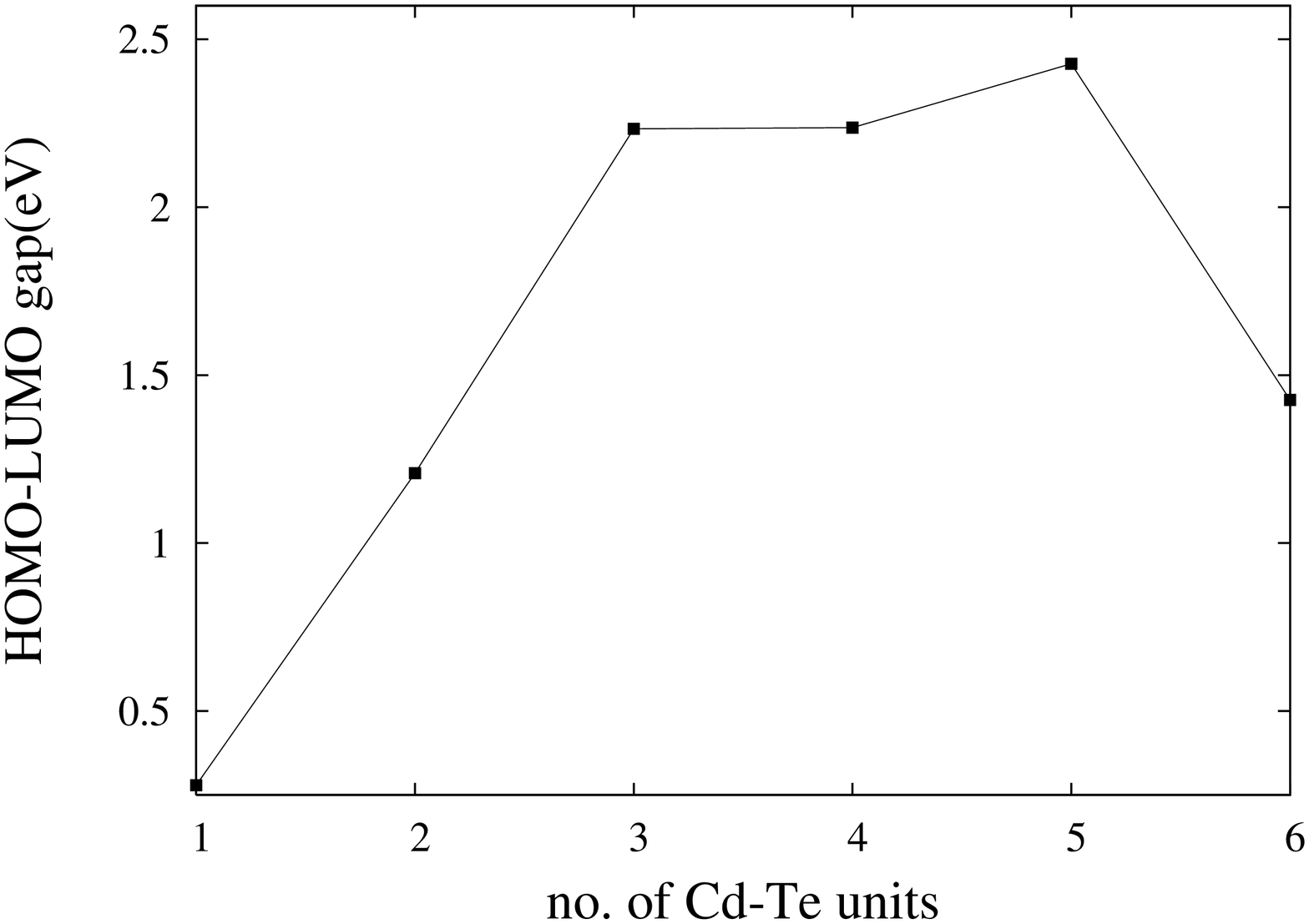}
\caption{HOMO-LUMO gap in eV {\it vs} number of Cd-Te units.}\label{bg} 
\end{figure}

\begin{figure}[h]
\centering
\includegraphics[height=8.5cm,angle=270] {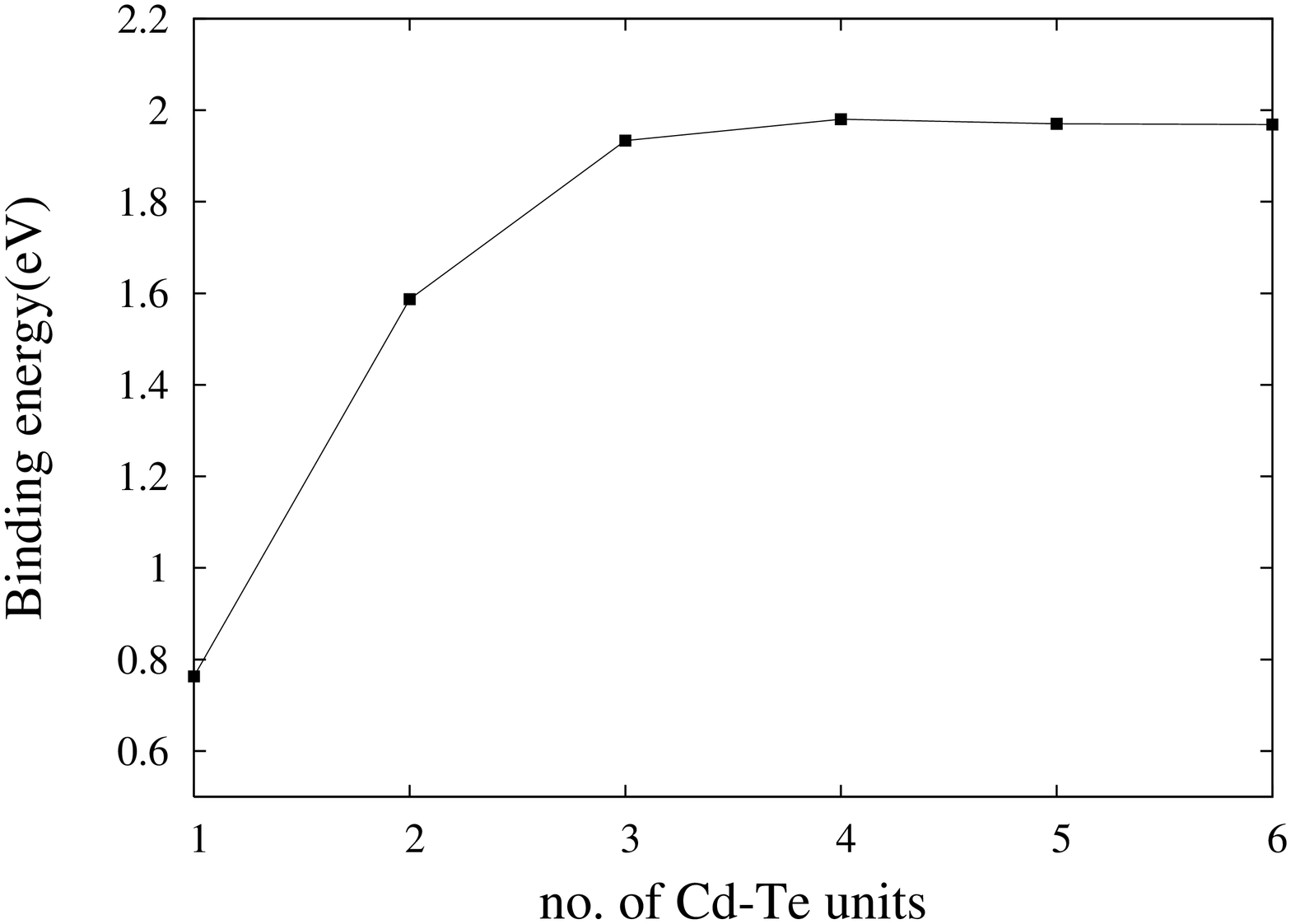}
\caption{Binding energy in eV {\it vs} number of Cd-Te units.}\label{be}
\end{figure}

The corresponding two curves for CdS and CdSe show similar trends~\cite{r6} and
the authors claim to infer the stability of clusters from the maximum in these
curves. However, we do not see any such correlation
for the clusters studied in the present work. 
It is seen that Cd$_5$Te$_5$ is more stable compared to its neighbours. It may
be mentioned that the number of electrons in this system are 90 which matches 
with the shell closing number for jellium-like systems.

We have calculated the vertical detachment energy (VDE), shown in Fig. \ref{ip},
and the electron affinity (EA), shown in Fig. \ref{ea}, using Eqs. (\ref{VDE})
and (\ref{EA}) by performing two self-consistent calculations for different 
elecronic systems without allowing any change in the lowest energy geometry of 
the neutral cluster.

\vspace{-0.1in}
\begin{equation}
VDE(Cd_nTe_n) = E({Cd_nTe_n}^+) - E(Cd_nTe_n)  \label{VDE}
\end{equation}

\vspace{-0.15in}
\begin{equation}
EA(Cd_nTe_n) = E({Cd_nTe_n}^-) - E(Cd_nTe_n)  \label{EA}
\end{equation}

\begin{figure}[h]
\centering
\includegraphics[height=8.5cm,angle=270] {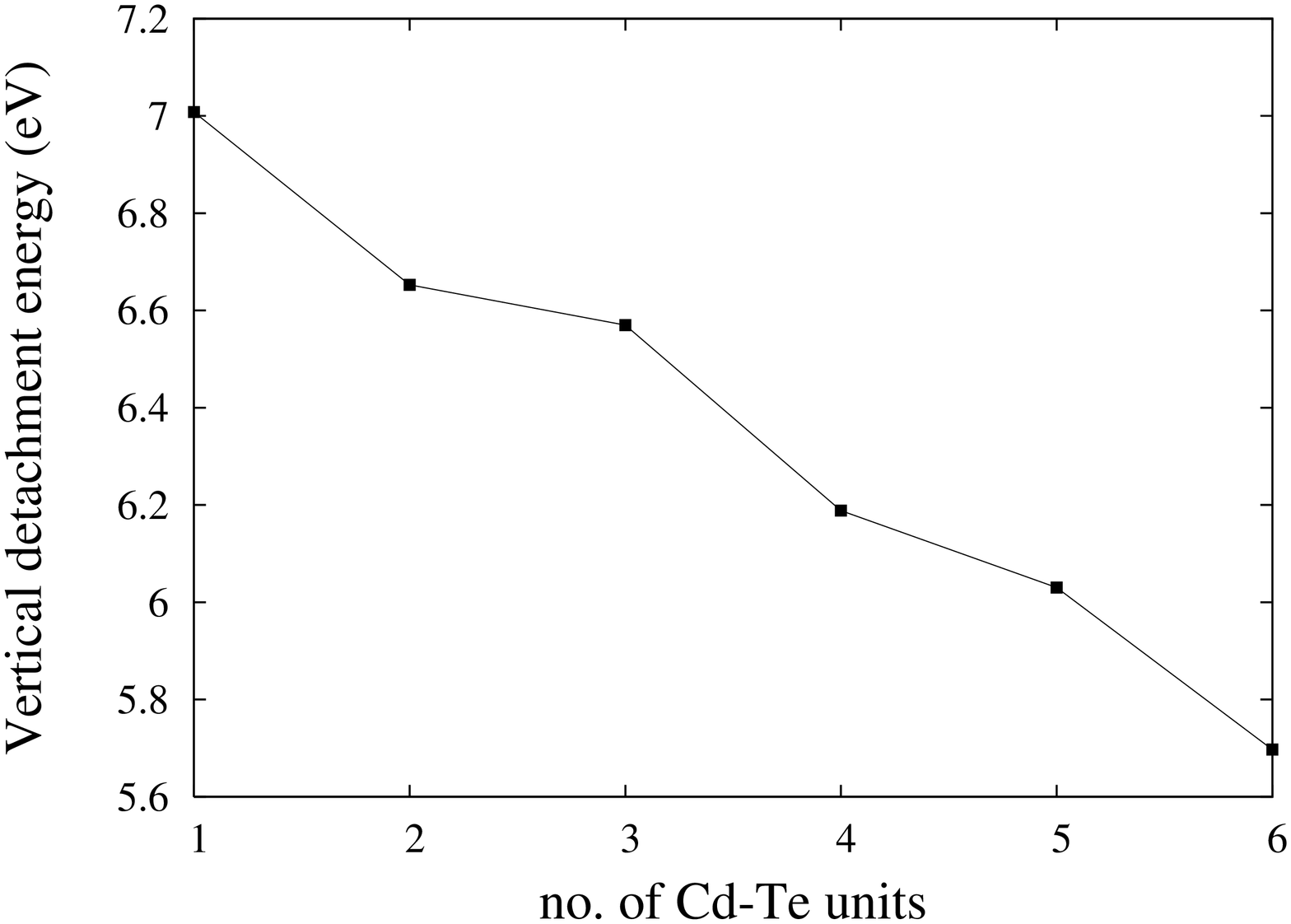}
\caption{Vertical detachment energy (eV) {\it vs} number of Cd-Te units.}
\label{ip}
\end{figure}

\begin{figure}[h]
\centering
\includegraphics[height=8.5cm,angle=270] {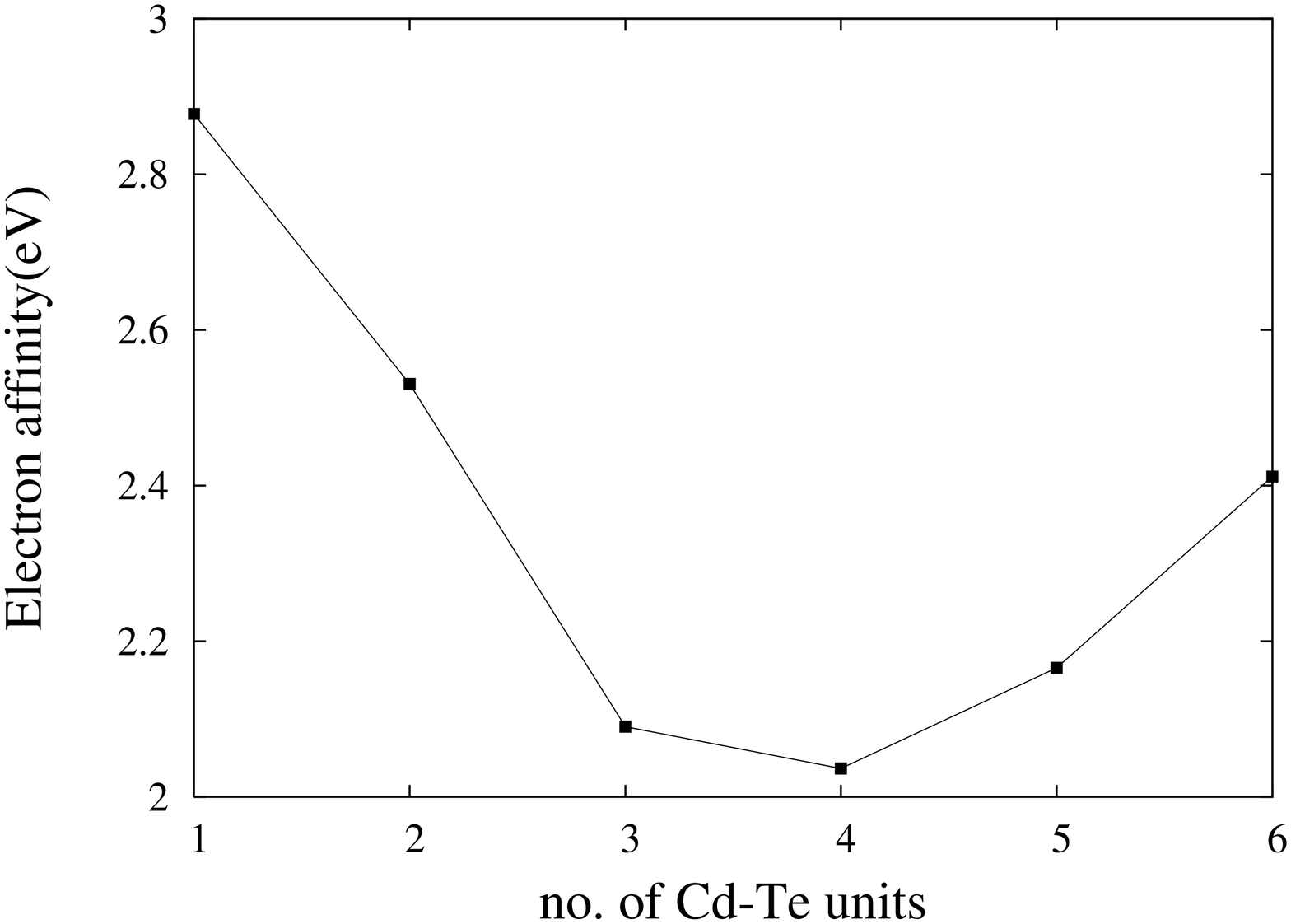}
\caption{Electron affinity in eV {\it vs} number of Cd-Te units.}\label{ea}
\end{figure}

In bulk CdTe, the bonding is largely covalent as the atomic number and hence 
the atomic radii for Cd and Te are very close. As a result, the Coulombic 
forces between the two species are balanced very well.
However due to difference in electronegativity of Cd and Te, the bonding
has some ionic character even in the bulk.

\begin{figure}[h]
\centering
\includegraphics[height=4.5cm,width=6.0cm,angle=180] {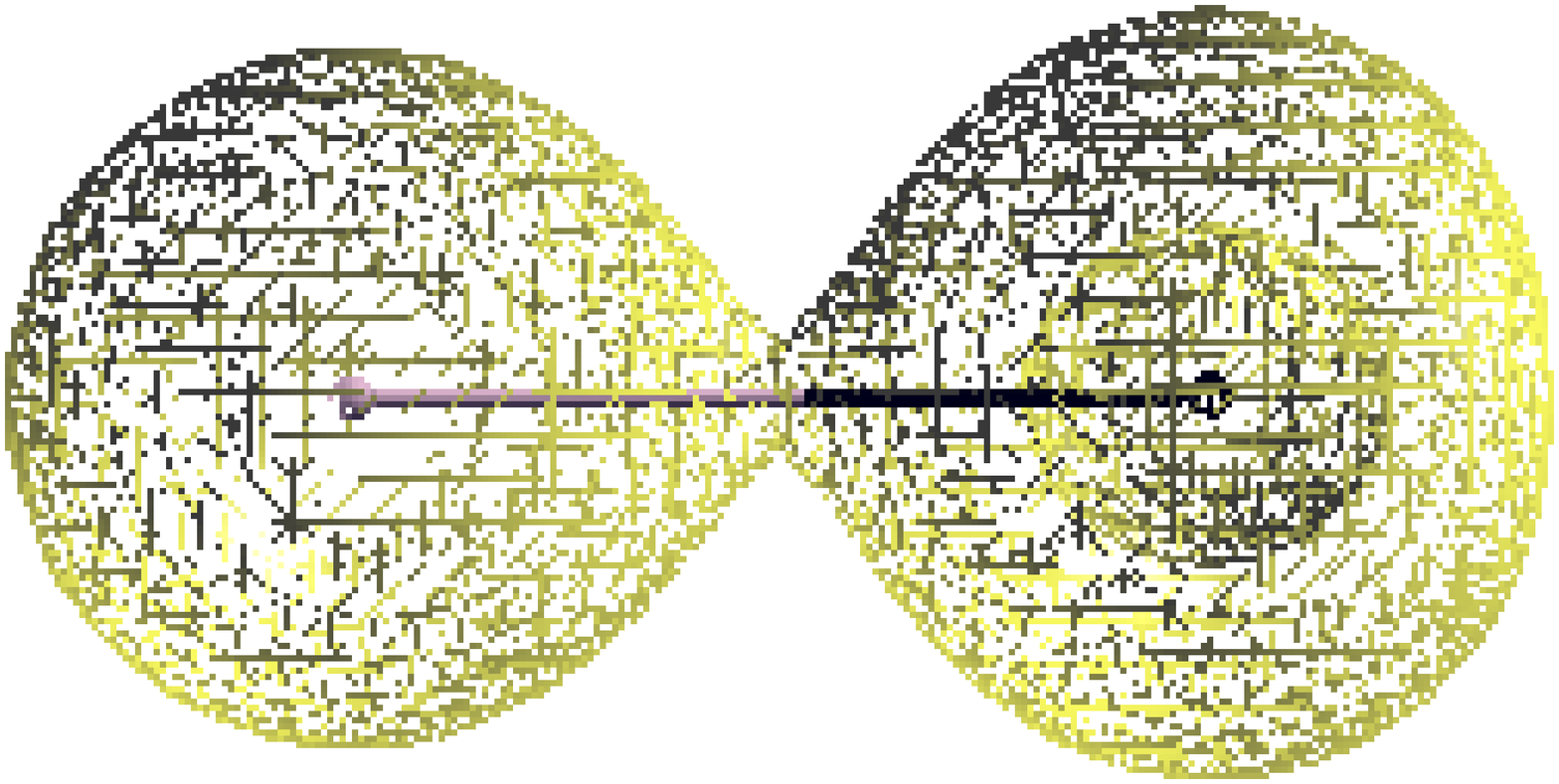}
\caption{Charge density iso-surface for CdTe dimer.}\label{cd1}
\end{figure}

\begin{figure}[h]
\centering
\includegraphics[height=4.5cm,width=6.0cm, angle=180] {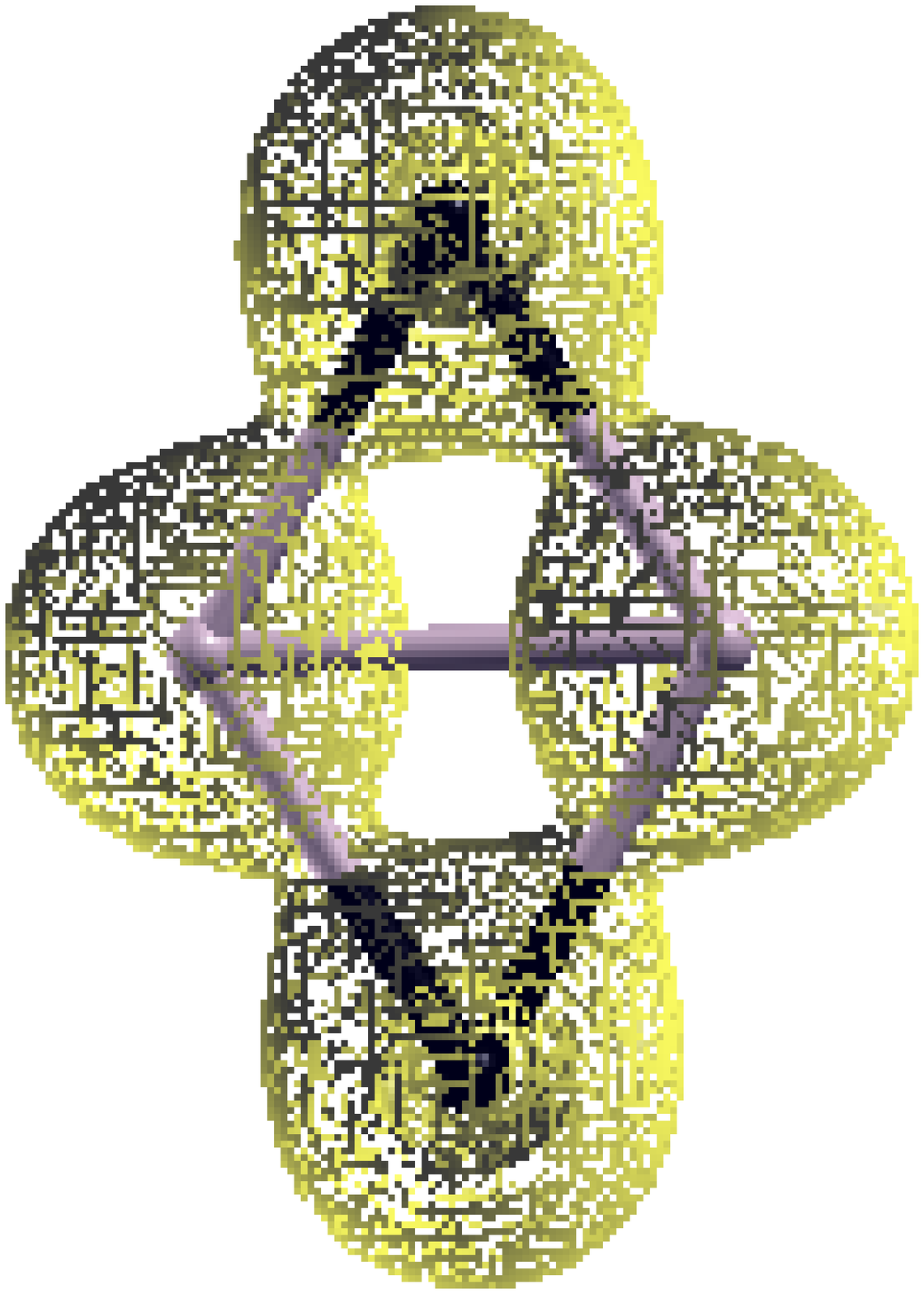}
\caption{Charge density iso-surface for Cd$_2$Te$_2$. The grey atoms are Cd and 
black atoms are Te.}\label{cd2}
\end{figure}

We show total charge density iso-surface plots for CdTe dimer and 
Cd$_2$Te$_2$ cluster
in Figs. \ref{cd1} and \ref{cd2} respectively, revealing covalent 
bonding. Similar charge density distributions are seen for other clusters too. 
No Cd-Cd or Te-Te near neighbour bonds are seen to be formed for clusters with 
$n > 2$.
The charge density plots
for CdS and CdSe~\cite{r6} show that the bonding has some ionic character and
is not purely covalent. 
Ionicity in bonding is more in CdS than in CdSe because of smaller atomic
 radius of sulphur. Ionicity is expected to be still smaller 
in CdTe.

An analysis of charge inside atomic spheres with covalent radii
shows that the distribution of charge in the Cd and Te spheres 
is almost same for planar structures for $n = 3-5$. There is more charge
 localization in planar structures in comparision to CdTe dimer, 
Cd$_2$Te$_2$ and the 3D structures.

A more detailed information regarding the bonding in these clusters 
can be obtained from the partial charge density plots of the molecular 
orbitals (MOs) in these clusters.

\begin{figure}[h]
\centering
\includegraphics[height=4.5cm,width=5.0cm,angle=360] {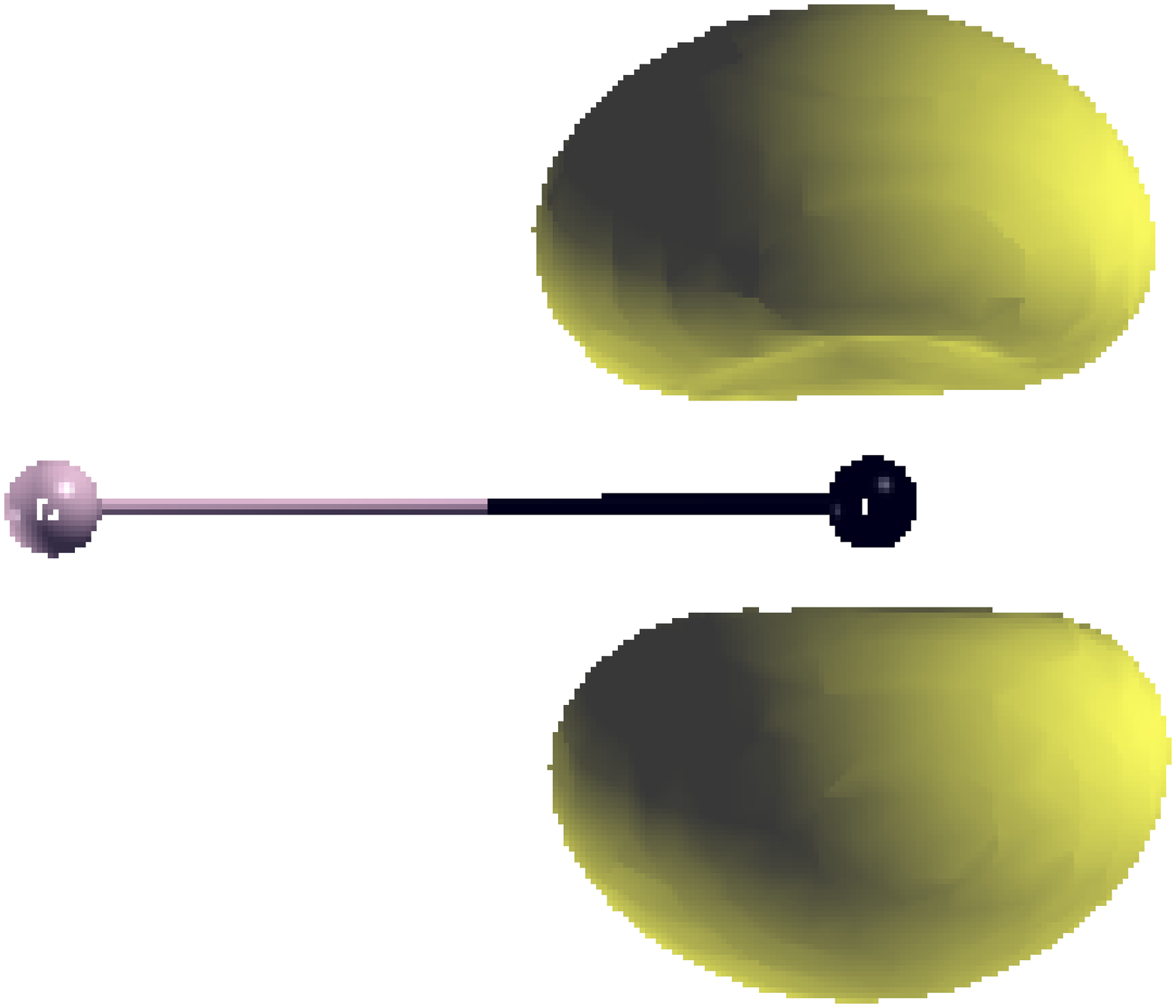}
\caption{Iso-surface of partial charge density of HOMO for CdTe dimer.}\label{homodimer}
\end{figure}

\begin{figure}[h]
\centering
\includegraphics[height=4.5cm,width=5.0cm,angle=360] {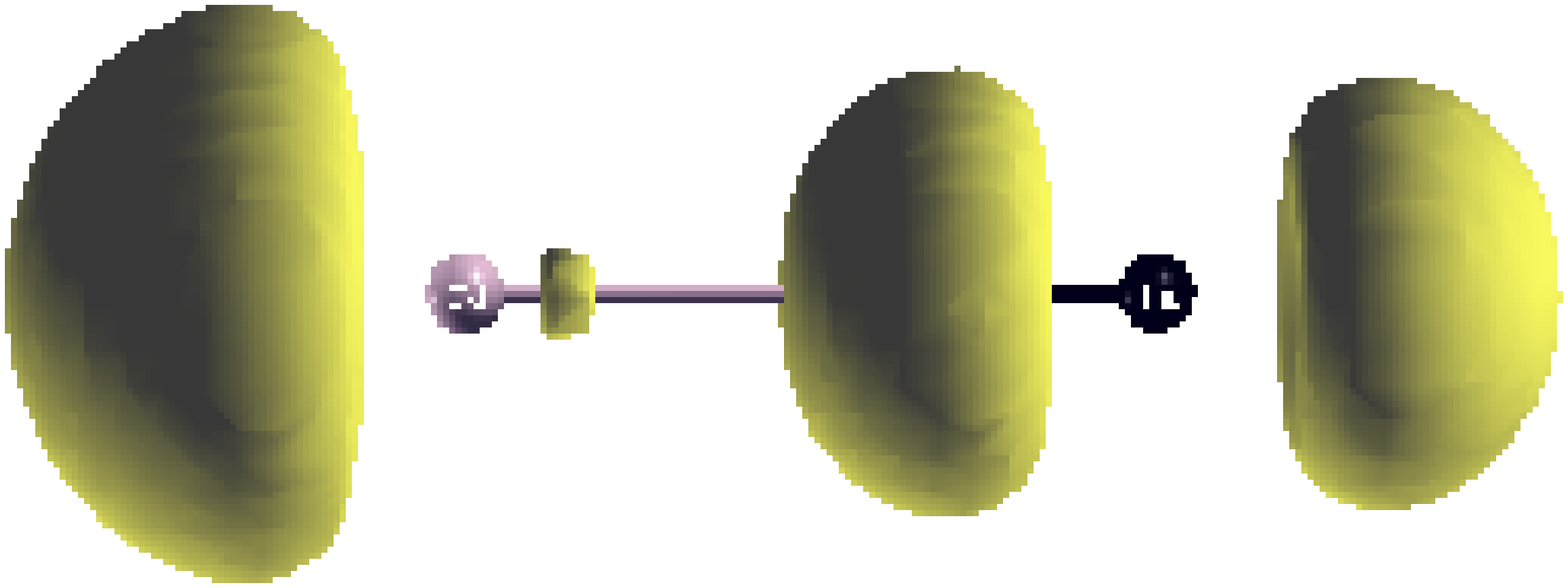}
\caption{Iso-surface of partial charge density of LUMO for CdTe dimer.}\label{lumodimer}
\end{figure}

From Figs. \ref{homodimer} and \ref{lumodimer}
 it is evident that the HOMO consists of purely Te $p$ and in the
LUMO Cd $s$ and $p$ hybridize to give a $sp$-hybrid orbital which in turn 
forms a $\sigma$ bond with the Te $p$ orbital.
Similar trends are seen for the plots for HOMO and LUMO of Cd$_3$Te$_3$
as shown in Figs. \ref{homocd3te3} and \ref{lumocd3te3}. 
Figure \ref{lumocd3te3} also shows the 
delocalization of the charge density in the central region of the clusters
indicating a semi-metallic character in the bonding. Thus, these systems 
will favour electron addition as is also evident from  Fig. \ref{ea}.
For other clusters the partial charge density for HOMO and LUMO have similar
nature.

\begin{figure}[h]
\centering
\includegraphics[height=4.5cm,width=5.0cm,angle=360] {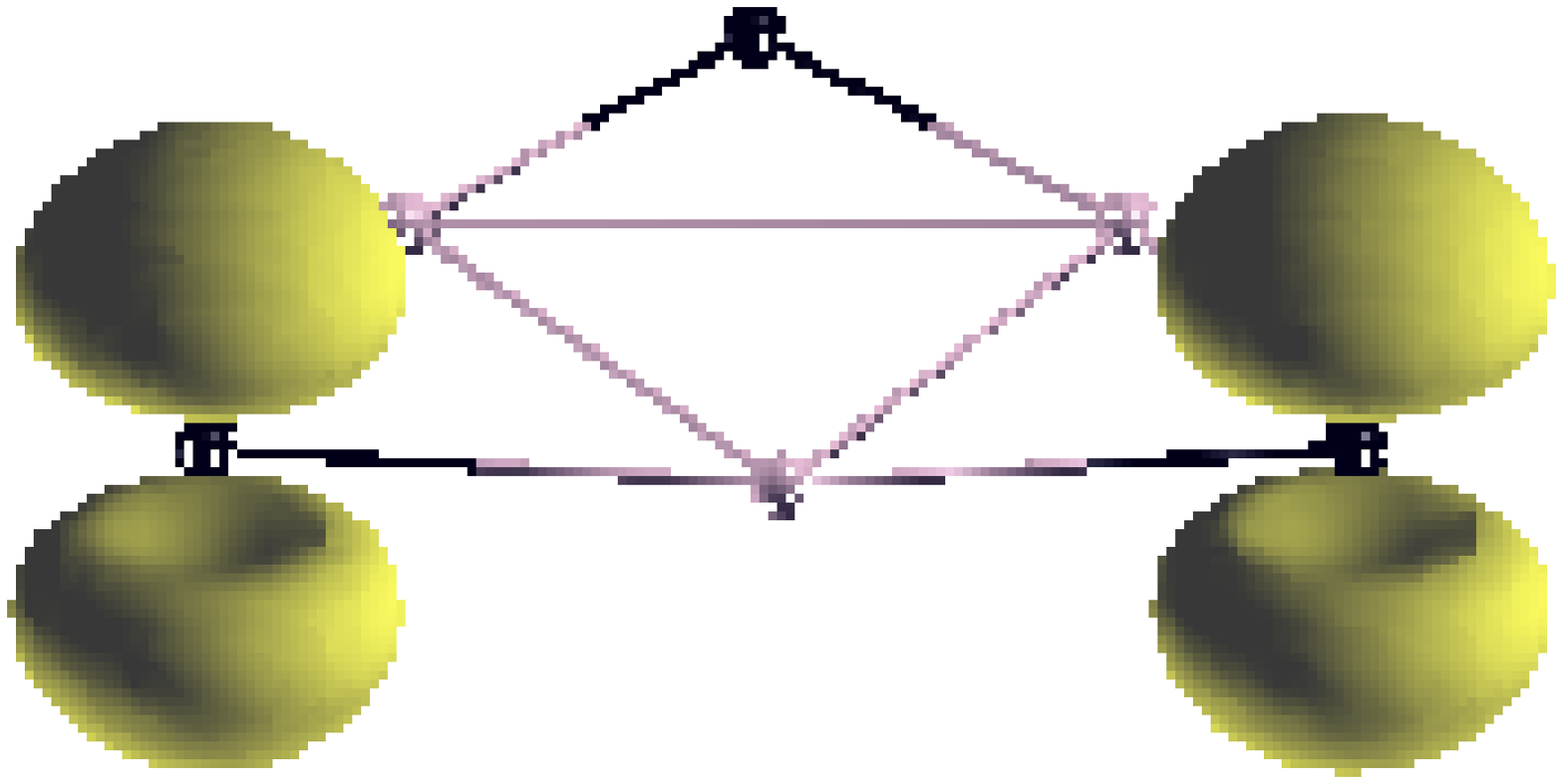}
\caption{Iso-surface of partial charge density of HOMO for Cd$_3$Te$_3$.}\label{homocd3te3}
\end{figure}

\begin{figure}[h]
\centering
\includegraphics[height=4.5cm,width=5.0cm,angle=360] {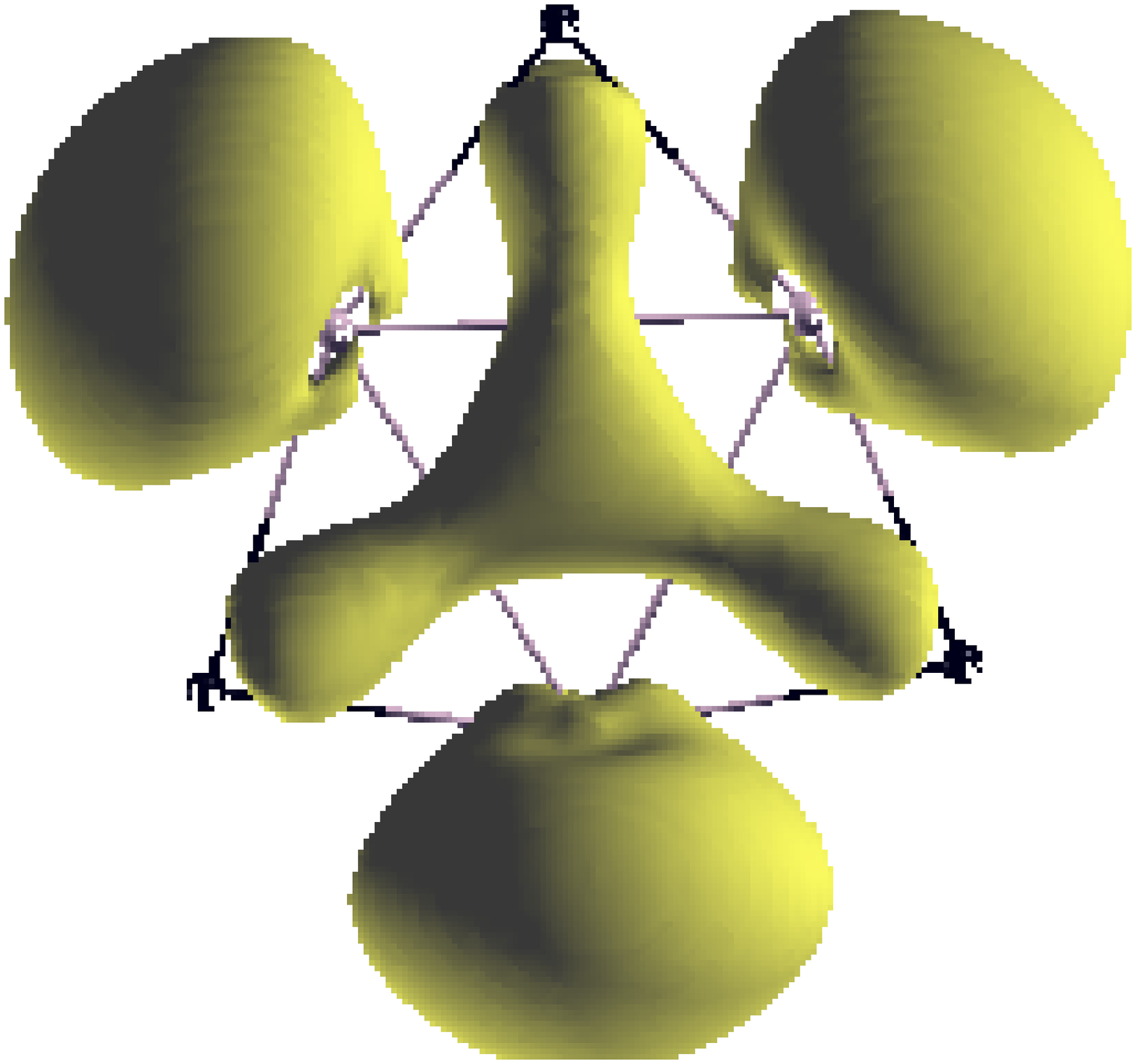}
\caption{Iso-surface of partial charge density of LUMO for Cd$_3$Te$_3$.}\label{lumocd3te3}
\end{figure}

As seen from the MO charge distribution of HOMO as one goes
from $n = 2$ to $n = 6$, an enhancement in delocalisation results
accounting for a steady decrease in the VDE as shown in Fig. \ref{ip}. 

\begin{figure}[h]
\centering
\includegraphics[height=7.5cm,width=8.0cm,angle=270] {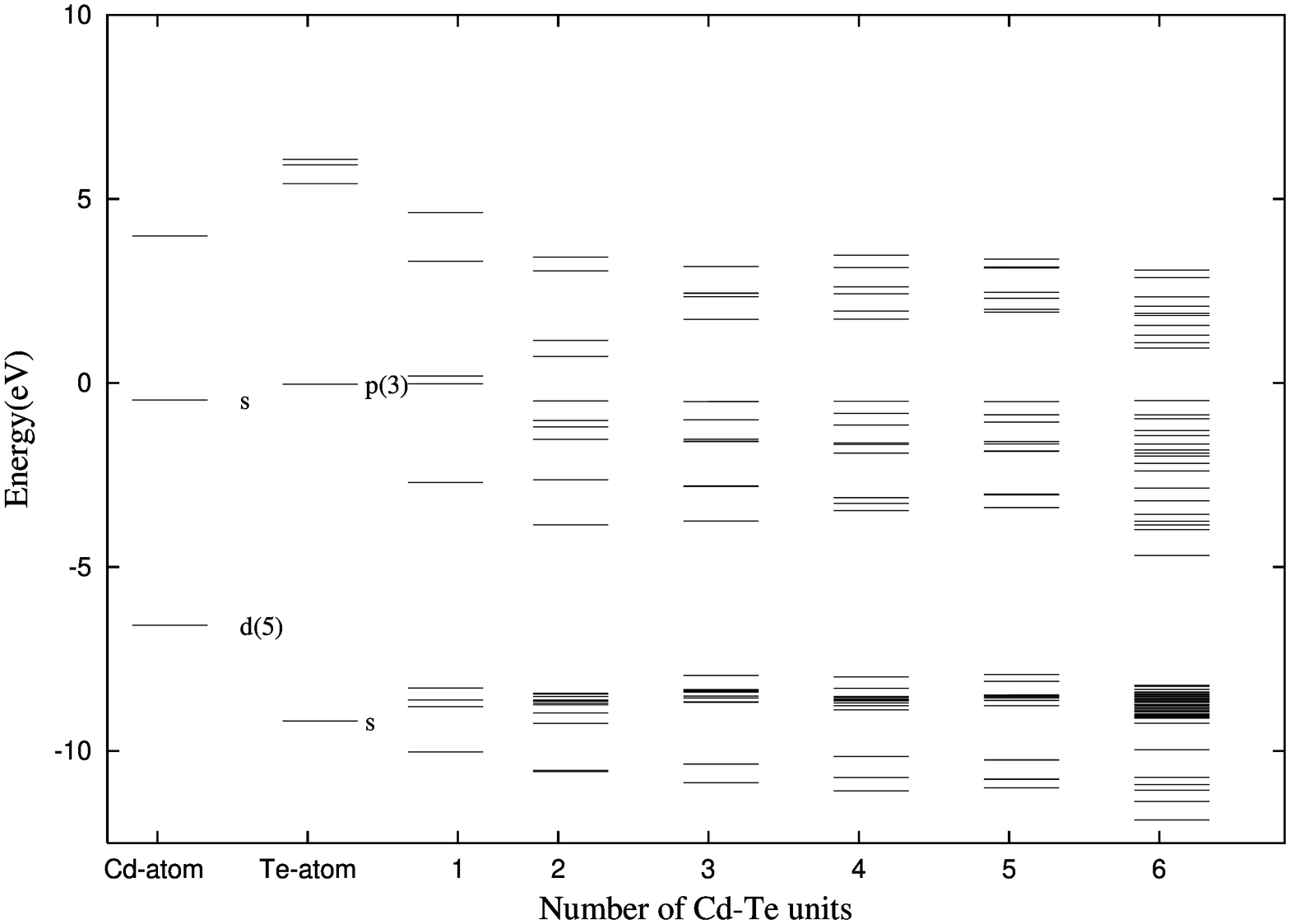}
\caption{ Energy level diagram of CdTe clusters.}\label{ene}
\end{figure}

In Fig. \ref{ene} we show the one-electron energies for all the CdTe clusters 
and the atomic levels of Cd and Te. The 
Cd $s$ and the Te $p$ levels lie in the same energy window in the atomic case.
Consequently, an intermixing of these orbitals occurs in the clusters.
A fat band analysis of the CdTe bulk band structure
(not shown here) using 
full potential linear augmented plane wave method~\cite{wien} reveals 
similar intermixing of orbitals. Thus, the bulk type hybridization of the 
orbitals is retained in clusters.
The movement of the the HOMO and LUMO in these small clusters is quite 
interesting. As the 
cluster size increases, the HOMO moves downward whereas the LUMO moves 
upward resulting in widening of the HOMO-LUMO gap. However, this shift is not 
uniform.
The HOMO moves downward slowly as compared to the LUMO which
moves upward rapidly. But for Cd$_6$Te$_6$, the HOMO shifts slightly 
upward and the LUMO shifts downward resulting in a smaller HOMO-LUMO gap 
as a consequence of planar to 3D transition in geometry.
\end{subsection}

\begin{subsection}{Properties of Cd$_m$Te$_n$ clusters}
CdTe, with many other II-VI semiconductors, in bulk has  ZB and wurtzite (W) 
phases possible.
There are also reports of size induced phase transitions
in ZnS~\cite{nan} and CdS~\cite{cds} nanoparticles. The first coordination
for central atom is identical in ZB and W phases. If we consider the
trigonal axis [111] in ZB and [001] in W then three near neighbours
are identical.

We have, therefore, performed calculations for four
larger non-stoichiometric CdTe clusters {\it viz.}
Cd$_{13}$Te$_{16}$ (central atom + 3 neighbouring shells), 
Cd$_{16}$Te$_{19}$ (central atom + 4 shells), Cd$_{16}$Te$_{13}$ and
Cd$_{19}$Te$_{16}$. The later two geometries are obtained from the former
two by interchanging the positions of Cd and Te atoms.
These clusters have been considered as fragments of the bulk
and hence their initial symmetry is T{$_d$} as shown in Figs.~\ref{side}(a) 
and \ref{top}(a).

\begin{figure*} 
\epsfxsize 2.2in
\epsffile{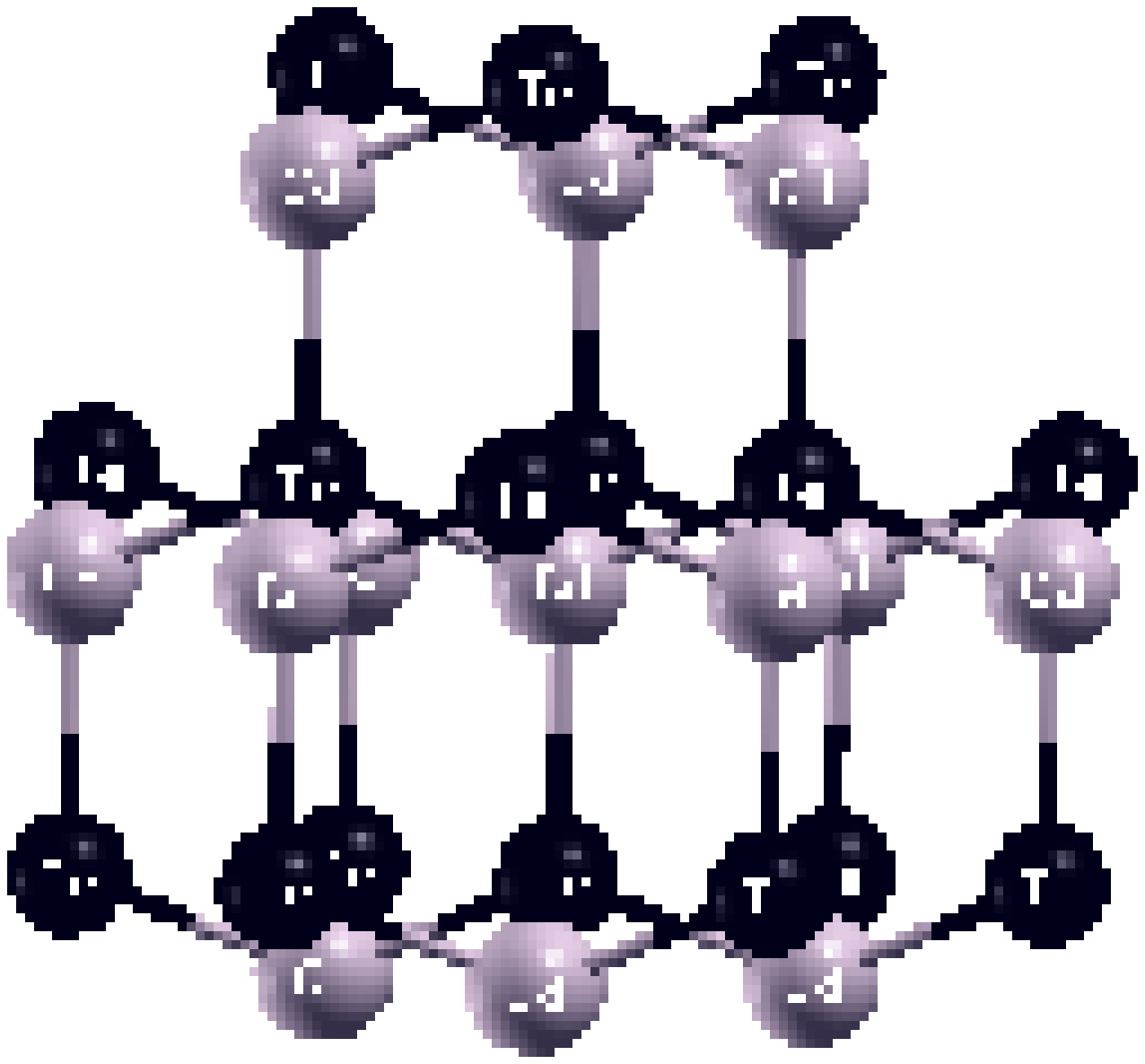}
\hspace{-4.5cm}
{(a) $Cd_{13}Te_{16}$ : initial}
\hspace{0.3cm}
\epsfxsize 2.2in
\epsffile{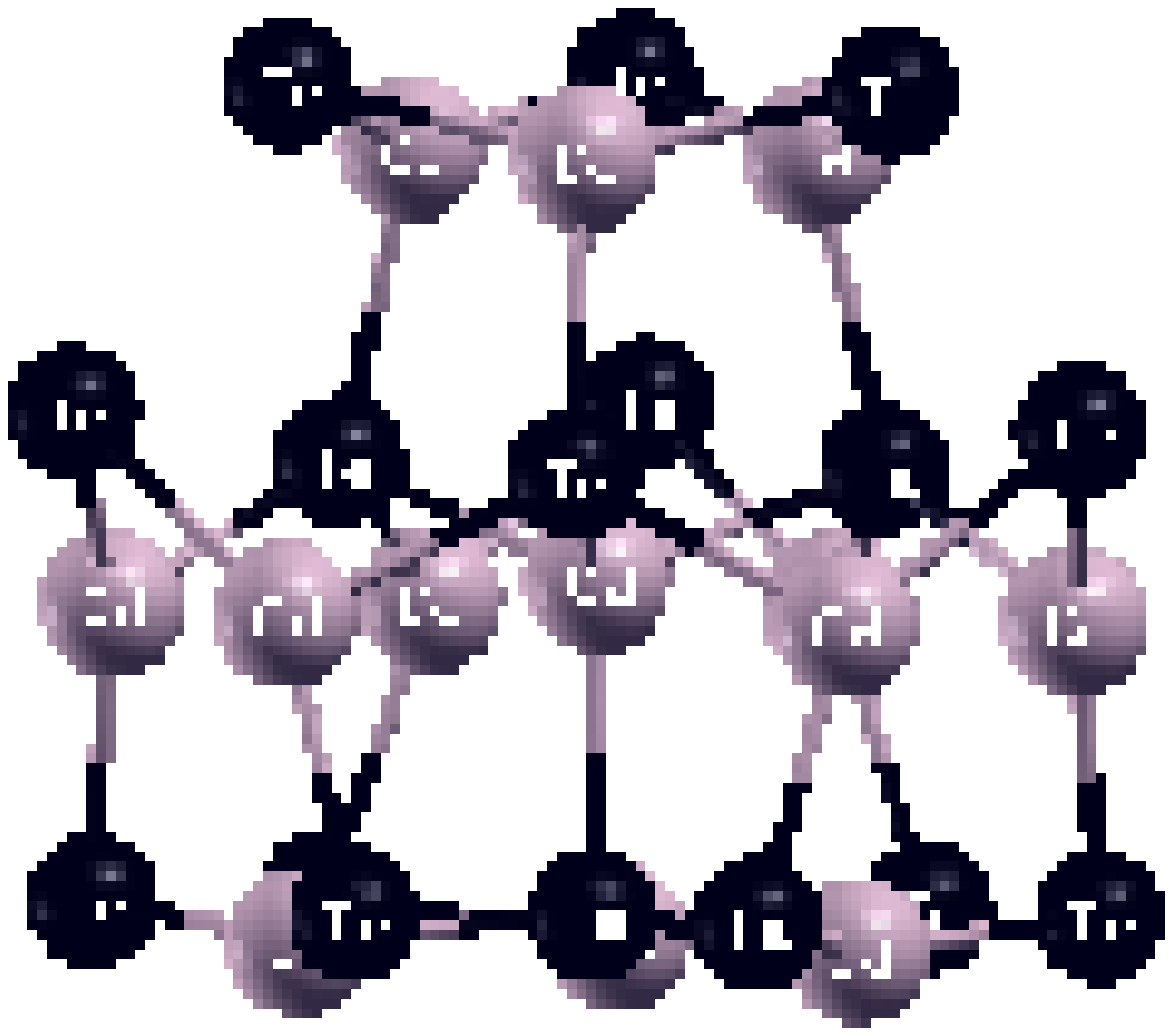}
\hspace{-4.5cm}
{(b) $Cd_{13}Te_{16}$ : relaxed}
\hspace{0.3cm}
\epsfxsize 2.2in
\epsffile{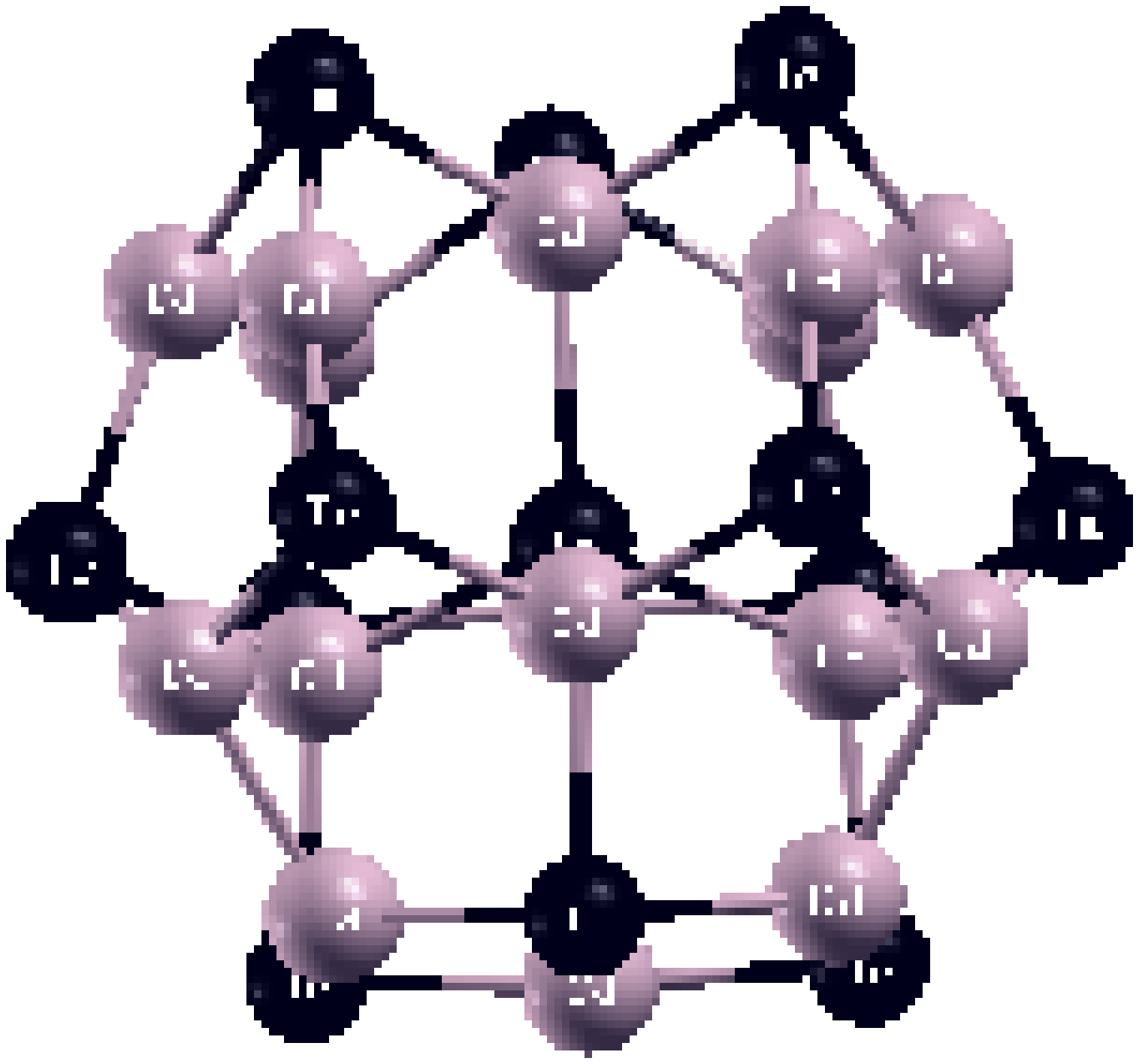}
\hspace{-4.5cm}
{(c) $Cd_{16}Te_{13}$ : relaxed}
\caption{Side view of the initial and relaxed geometries.}\label{side}
\end{figure*}

Cd$_{13}$Te$_{16}$ has a central Cd atom and the side view of its relaxed 
structure is shown in Fig.~\ref{side}(b). Figure~\ref{side}(c) shows the 
side view of relaxed structure of Cd$_{16}$Te$_{13}$ which 
has a central Te atom.
During relaxation the central Te atom pushes the neighboring
Cd atoms outward. 
Figure \ref{top} shows the top views of initial geometry for 
Cd$_{16}$Te$_{19}$(a) and
the relaxed geometries for Cd$_{16}$Te$_{19}$(b) and
Cd$_{19}$Te$_{16}$(c).
It is found that upon relaxation only the Te-rich clusters retain their 
T{$_{d}$} symmetry whereas 
the Cd-rich clusters attain lower symmetry structures.

\begin{figure*} 
\epsfxsize 2.2in
\epsffile{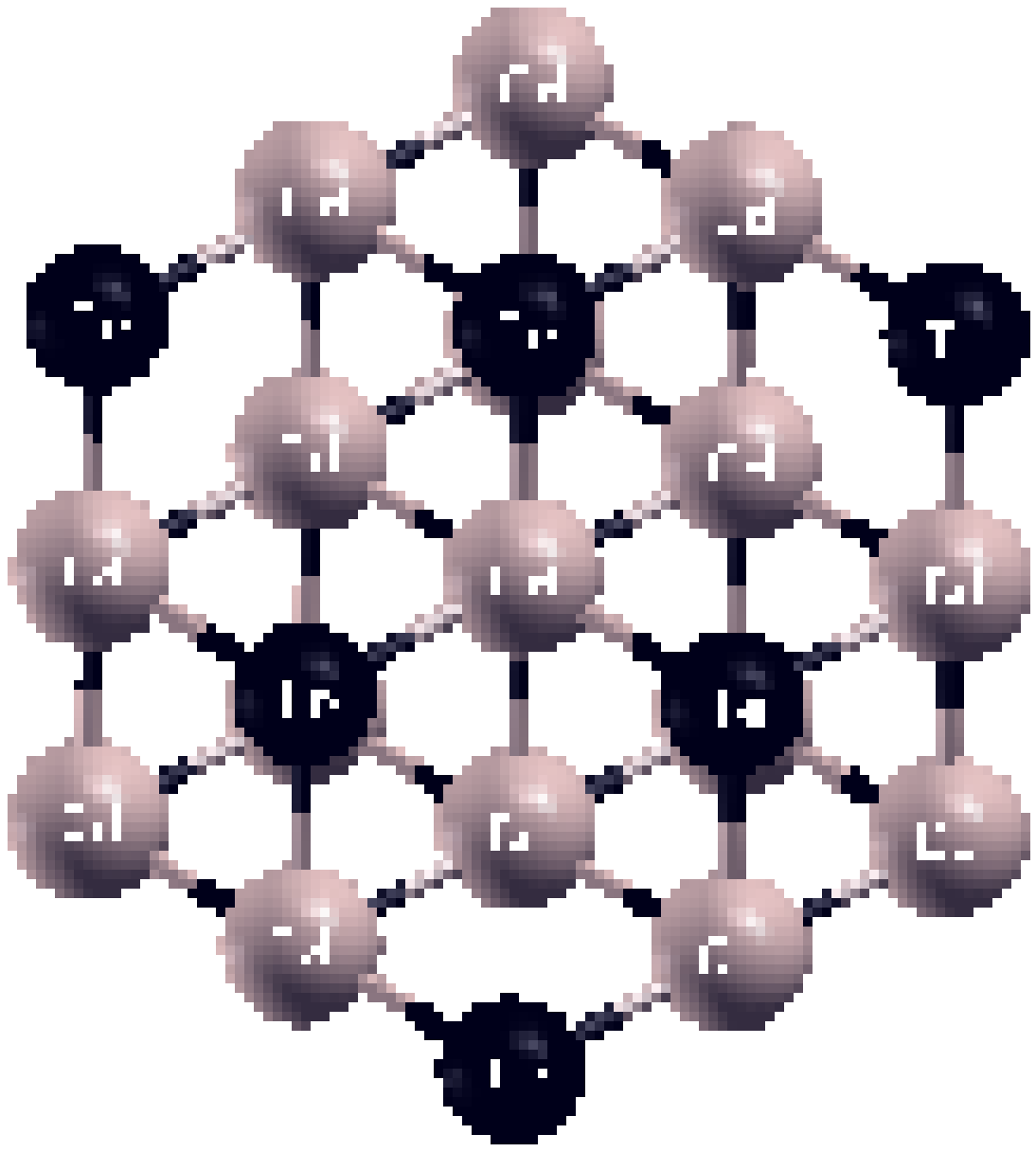}
\hspace{-4.5cm}
{(a) $Cd_{16}Te_{19}$ : initial}
\hspace{0.3cm}
\epsfxsize 2.2in
\epsffile{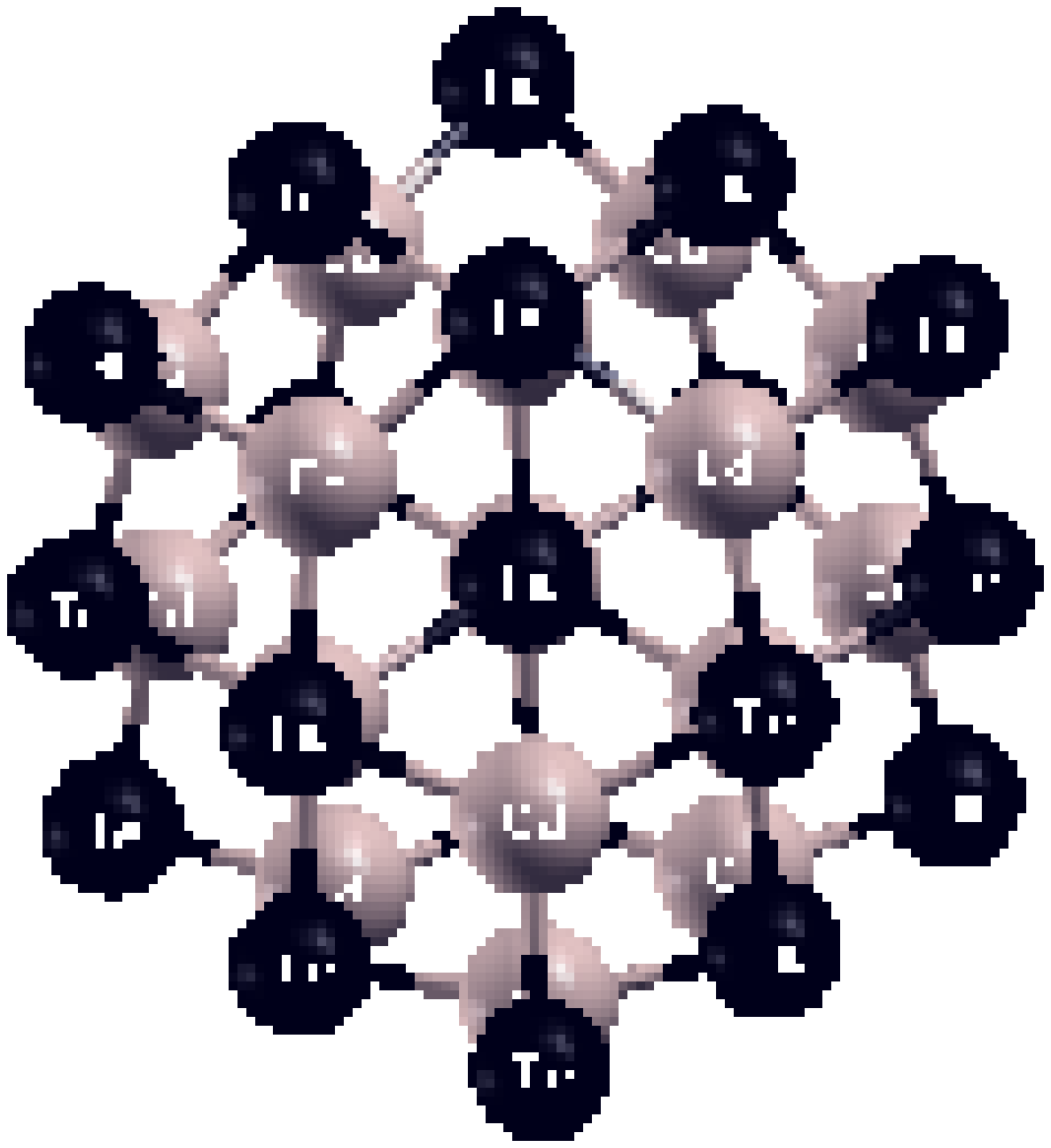}
\hspace{-4.5cm}
{(b) $Cd_{16}Te_{19}$ : relaxed}
\hspace{0.3cm}
\epsfxsize 2.2in
\epsffile{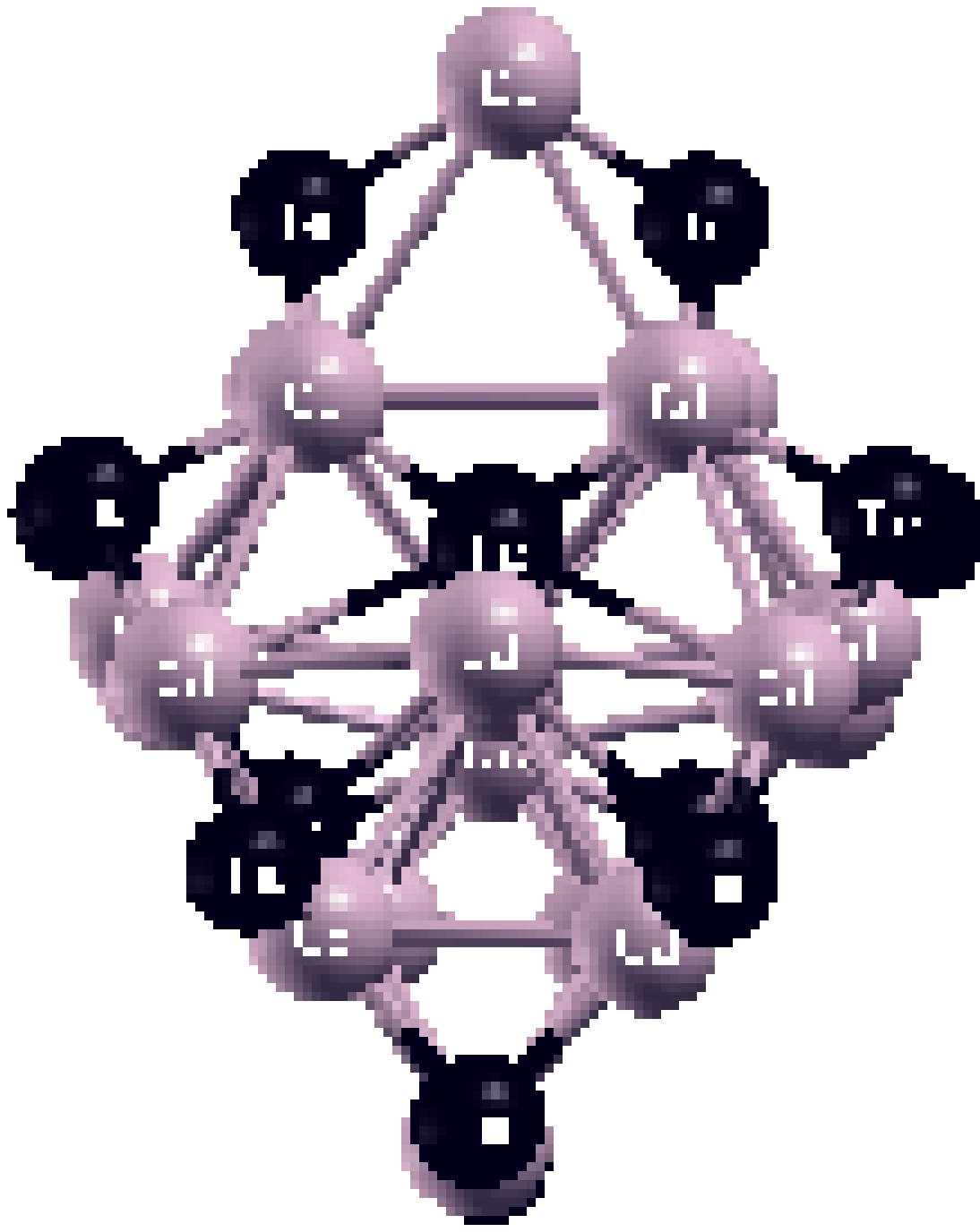}
\hspace{-4.5cm}
{(c) $Cd_{19}Te_{16}$ : relaxed}
\caption{Top view of the initial and relaxed geometries.}\label{top}
\end{figure*}

We summarize in table \ref{T1} some of the structural and electronic properites
of Cd$_m$Te$_n$ clusters. 
It may be mentioned that the binding
energy of CdTe dimer is  0.76 eV/atom whereas the
cohesive energy of CdTe bulk is 4.9 eV.

\begin{table}[h]
\caption{Summary of Structural and Electronic properties of
relaxed Cd$_m$Te$_n$($m{\neq}n$) clusters.
PG is the observed point group symmetry, BE is binding energy expressed in 
(eV/atom), $l$ is average Cd-Te bond length in $\AA$, $\theta$ is the Y-X-Y
bond angle at the central atom X of the cluster and E$_g$ is the HOMO-LUMO
gap in eV.}\label{T1} 
\begin{ruledtabular}
\begin{tabular}{cccccc} 
\hline
Clusters & PG & BE & $l$  & $\theta$ & E$_g$ \\
\hline
Cd$_{13}$Te$_{16}$ & T{$_d$} & 2.13 & 2.81 &  109.47$^{\circ}$ & 0.195 \\
\hline
Cd$_{16}$Te$_{13}$ & C{$_{3v}$} & 1.95 & 2.84 & 109.48$^{\circ}$ & 0.659  \\
\hline
Cd$_{16}$Te$_{19}$ & T{$_{d}$} & 2.14 & 2.82 & 109.47$^{\circ}$ & 0.088 \\
\hline
Cd$_{19}$Te$_{16}$ & C$_1$ & 1.95 & 2.81 & 91.31$^{\circ}$  & 0.964 \\
\hline
\end{tabular}
\end{ruledtabular}
\end{table}

A comparison of the binding energies of the above
clusters indicates that
Cd$_{13}$Te$_{16}$ and Cd$_{16}$Te$_{19}$ are
more stable than Cd$_{16}$Te$_{13}$ and Cd$_{19}$Te$_{16}$. In the 
former group the Te atoms are on the surface and hence these structures are
energetically favoured as discussed earlier for smaller clusters.
For the Te-rich clusters, the relaxation is not strong enough to break their 
initial T$_d$ symmetry.
On the other hand, the Cd-rich clusters, being weakly 
bound, go to lower 
symmetry structures upon relaxation. A comparision of binding energy per atom
for Cd$_m$Te$_n$ with corresponding Hg$_m$Te$_n$ clusters~\cite{r4} shows an
increase.
This is attributed to symmetry allowed $pd$ interactions. The $d$ levels
of Hg are closer than those of Cd to the $p$ levels of Te. 
Therefore $pd$-repulsion 
induced bond weakening is more in HgTe than in CdTe clusters resulting in a 
reduction of binding energy.

A very interesting feature observed in the Cd-rich clusters
is Jahn-Teller distortion. The initial HOMO-LUMO gap 
for Cd$_{16}$Te$_{13}$ and Cd$_{19}$Te$_{16}$ 
is nearly zero. But upon relaxation, they attain an 
appreciable HOMO-LUMO gap. This is a clear indication that
the energy levels which are nearly degenerate in the initial
structure have moved apart during relaxation. Thus, relaxation has
destroyed symmetry resulting in the lifting of degeneracy. On the other
hand for the Te-rich clusters, the HOMO-LUMO are not degenerate
(though close to each other) and thus there is no Jahn-Teller
distortion.
Similar trend has also been observed in non-stoichiometric 
HgTe clusters~\cite{r4}.
The Hg-rich clusters upon relaxation lose their initial T$_d$
symmetry. As a result, they attain a definite bandgap and
a semi-metal to semiconductor transition is observed.
Dalpian {\it et al.} have shown that
symmetry is an important parameter in deciding the properties of 
nanoclusters~\cite{dalpian}.

\end{subsection}
\end{section}

\begin{section}{Conclusion}
The ground state geometries of Cd$_n$Te$_n$ ($1{\leq}n{\leq}6$) 
were calculated using {\it ab-initio} DFT. The structures do not resemble that 
of the bulk phase. The Cd atoms tend to form a core group surrounded by the 
chalcogenide atoms because of the Coulomb repulsion of the
lone pair of $p$-electrons on Te. For Cd$_n$Te$_n$,
(1${\leq}$n${\leq}$5), the planar geometry is the lowest
energy configuration as Te-Cd-Te prefer to 
form linear bond. But as the number of atoms increases further, higher 
coordination takes over and three dimensional structures result.
Bonding in CdTe clusters 
is mostly covalent in nature with partial ionic character.
The degree of ionicity changes with $n$.
Hybridization of orbitals in these clusters is observed to be 
same as that in the bulk.  

The relaxed structures of non-stoichiometric clusters, Cd$_m$Te$_n$
with ($m,n = 13,16,19$ and $m \neq n$),
were obtained using {\it ab-initio} DFT with the wavefunction projected 
to real space. The Te-rich clusters with Te atoms on the surface,
 are found to be more stable than the 
Cd-rich clusters. The symmetry of the cluster plays
 a vital role in determining its HOMO-LUMO gap. It was found that 
the Cd-rich cluster lose their initial T$_{d}$ symmetry and attain 
a large HOMO-LUMO gap as compared to the Te-rich clusters.
\end{section}

\begin{section}{Acknowledgment}
We acknowledge financial support from the Department of Science \& Technology, 
Governement of India. We also thank IUCAA, Pune and Centre for Modeling and
Simulation, University of Pune for use of their
computational facilities.
\end{section}

\end{document}